\documentclass[nofootinbib,a4paper,12pt]{article}
\usepackage{jheppub}
\usepackage[T1]{fontenc}
\usepackage[utf8]{inputenc}
\usepackage{float}
\usepackage{slashed}
\usepackage{multicol,multirow}
\usepackage{amssymb}
\usepackage{amsmath}
\usepackage{subcaption}
\usepackage{array}
\usepackage{color}
\usepackage{hyperref}
\hypersetup{colorlinks=true}
\usepackage{longtable}
\usepackage{graphics,graphicx,epsfig,ulem,booktabs}
\numberwithin{equation}{section}

\newenvironment{longfigure}
  {\longtable}
  {\endlongtable}

\definecolor{blue}{rgb}{0.2, 0.4, 1.0}
\definecolor{green}{rgb}{0.1,0.8,0.2}
\definecolor{orange}{rgb}{0.95,0.45,0.0}
\definecolor{cyan}{rgb}{0.0,0.75,0.8}

\newcolumntype{C}[1]{>{\centering\let\newline\\\arraybackslash\hspace{0pt}}m{#1}}

\newcommand{\tud}{\tau (B^+)/\tau (B_d)}

\newcommand{\Oq}{{\cal O}}
\newcommand{\Oqp}{{\cal O}^{\prime}}
\newcommand{\Q}{{\cal Q}}

\newcommand{\B}{{\cal B}}
\newcommand{\Bp}{{\cal B}^\prime}
\newcommand{\BM}{{B}}

\title{
\begin{center}
\boldmath 
 Taming New Physics in $b \to c \bar u d (s) $ with $\tud$ and $a_{sl}^d$
\end{center}
}
\preprint{SI-HEP-2022-32}

\author[a]{Alexander Lenz,}
\author[a]{Jakob M\"uller,}
\author[a]{Maria Laura Piscopo,}
\author[a]{Aleksey V. Rusov}

\affiliation[a]{Physik Department, Universit\"{a}t Siegen, 
Walter-Flex-Str. 3, 57068 Siegen, Germany}

\emailAdd{alexander.lenz@uni-siegen.de}
\emailAdd{jakob2.mueller@student.uni-siegen.de}
\emailAdd{maria.piscopo@uni-siegen.de}
\emailAdd{rusov@physik.uni-siegen.de}

\abstract{
Inspired by the recently observed tensions between the experimental data and the theoretical predictions, based on QCD factorisation, for several colour-allowed non-leptonic $B$-meson decays,
we study the potential size of new physics (NP) effects in the decay channels $b \to c \bar{u} d(s)$.
Starting from the most general effective Hamiltonian describing the $b \to c \bar{u} d(s)$ transitions,
we compute NP contributions to the theoretical predictions of $B$-meson lifetime and of $B$-mixing observables.
The well-known lifetime ratio $\tud$ and the experimental bound on the semi-leptonic CP asymmetry $a_{sl}^d$, provide strong, complementary constraints on some of the NP Wilson coefficients.
}

\begin{document}

\maketitle

\section{Introduction}
Recent theoretical studies~\cite{Huber:2016xod,Bordone:2020gao,Cai:2021mlt,Endo:2021ifc,Beneke:2021jhp,Fleischer:2021cct,Fleischer:2021cwb} indicate a discrepancy between the measured rates for non-leptonic 
$B$-meson decays triggered by the $b \to c \bar{u} d (s)$ transitions, e.g.~$\bar B_s \to D_s^+ \pi^-$~\cite{LHCb:2012wdi,LHCb:2013vfg,LHCb:2021qbv,Belle:2022afp} and their Standard Model (SM) predictions based on the QCD factorisation~(QCDF) framework~\cite{Beneke:2000ry}.
As $b \to c \bar{u} d$ is one of the two CKM leading non-leptonic $b$-quark decays, 
an explanation of this tension that goes beyond the SM (BSM) might seem at first sight unlikely, 
however at the moment this possibility can not be ruled out, 
see e.g.~Refs.~\cite{Cai:2021mlt, Bordone:2021cca}. Moreover, 
apart from the discrepancies in hadronic decays, 
BSM effects in non-leptonic tree-level $b$-quark decays have received some attention in the recent literature.
The presence of CP violating new contributions to $b \to c \bar u d $ transitions could lead to interesting
effects like the modification of the CKM
 angle~$\gamma$ by several degrees
 \cite{Brod:2014bfa}, while
 BSM effects in the channels $b \to c \bar{c} d$
 and $b \to c \bar{u} d$ could enhance the decay rate difference of the $B_d-\bar{B}_d$
 system \cite{Bobeth:2014rda} to an extent
 that the di-muon asymmetry measured by the
 D0 collaboration \cite{D0:2010sht} would be
 consistent with measurements of the 
 semileptonic CP asymmetries
 \cite{Amhis:2022mac}. 
 A general study of the allowed modification of the Wilson coefficients of the SM current-current operators, including also new CP violating effects, performed in Ref.~\cite{Lenz:2019lvd}, finds sizable space for new physics (NP) contributions in $b \to c \bar{u} d$ decays, particularly for the color-rearranged operator. 
Finally, BSM effects in tree-level $b \to c \bar{c}s $ transitions could also be responsible for
some of the observed $B$~anomalies~\cite{Jager:2017gal,Jager:2019bgk}. 
\\
While to better understand the nature of the current tensions in non-leptonic $B$-meson decays,
further investigations of unaccounted hadronic effects within 
the QCDF method would be highly desirable, in this paper, following a model-independent approach, 
we try to answer the question whether BSM effects in the $b \to c \bar{u} d(s)$ 
decay channels are consistent with other flavour observables 
\footnote{A clear experimental test of this assumption for the case of complex 
new physics couplings was suggested in Refs.~\cite{Gershon:2021pnc, Fleischer:2016dqd}.}. 
Specifically, extending the corresponding SM weak effective Hamiltonian 
with 20 NP operators with the most general Dirac structures, 
we study how the assumption of NP in $b \to c \bar u d(s)$ transitions would 
impact the theoretical predictions of the lifetime ratio $\tau(B^+)/\tau(B_d)$ 
and of the mixing observables $\Delta \Gamma_{d,s}$ and $a_{sl}^{d,s}$. 
Moreover, the effect of NP on colour-allowed, tree-level non-leptonic 
$B$-meson decays triggered by the $b \to c \bar u d (s)$ transitions,
like $\bar B_s \to D_s^+ \pi^-$, has been already studied in Ref.~\cite{Cai:2021mlt},
and in our numerical analysis we also compare our results with 
the ones presented in the latter reference.
\\
The lifetime ratio $\tud$ is by now experimentally determined with high precision~\cite{Amhis:2022mac}
\begin{equation}
   \frac{\tau (B^+)}{\tau (B_d)}^{\rm Exp.}
    = 1.076 \pm 0.004 \, .
\label{eq:ratio_Exp}
\end{equation}
On the theoretical side, predictions for this ratio can be obtained within the framework of the heavy quark expansion (HQE), which has proven to be a powerful method to perform systematic studies of inclusive decay widths of heavy hadrons, see e.g.\ the review~\cite{Lenz:2014jha}.
Based on the calculations in Refs.~\cite{King:2021jsq,King:2021xqp,Lenz:2020oce,Piscopo:2021ogu,Mannel:2020fts,Kirk:2017juj,Lenz:2013aua,Gabbiani:2004tp,Franco:2002fc,Beneke:2002rj}, within the SM,
the most recent value of the ratio reads~\cite{Lenz:2022rbq}
\begin{equation}
    \frac{\tau (B^+)}{\tau (B_d)}^{\rm HQE}
    =
    1.086 \pm 0.022 \, ,
\label{eq:ratio_HQE}
\end{equation}
in perfect agreement with the experimental measurements, albeit with much larger uncertainty.
In the presence of physics beyond the SM, new decay channels of the $b$-quark would also contribute to the total lifetime of the $B$
meson, and consequently modify the lifetime ratio according to 
 \begin{equation}
 \frac{\tau (B^+)}{\tau (B_d)}^{\rm HQE} \!\!\!\!\! = 1 + 
\left[\Gamma^{\rm SM} (B_d) - \Gamma^{\rm SM} (B^+) \right] 
\tau^{\rm Exp.} (B^+) 
+
\left[\Gamma^{\rm BSM} (B_d) - \Gamma^{\rm BSM} (B^+) \right]
\tau^{\rm Exp.} (B^+),
\label{eq:ratio_BSM}
\end{equation}
so that, by comparing with the corresponding experimental determination, Eq.~\eqref{eq:ratio_BSM} can be used to constrain the favoured parameter space for a specific set of BSM operators.
It is worth stressing already here that, while the $ b \to c \bar{u} d$ decay
represents the CKM leading contribution to $\tud$, the $ b \to c \bar{u} s$ channel is singly Cabibbo-suppressed and, at the current accuracy of our analysis, only enters the decay width of the $B^+$ meson but not that of the $B_d$. Hence, under the assumption of having the same BSM contribution in both
$b \to c \bar u d$ and $b \to c \bar u s$ transitions, our bounds on the parameter space from $\tud$ are dominated by NP in $b \to c \bar u d$, with $ b \to c \bar{u} s$ leading to only a small correction. 
We also emphasise that the lifetime ratio $\tud$ is particularly clean from a theoretical point of view, since in the limit of isospin symmetry only the contribution of four-quark operators appears on the r.h.s.\ of Eq.~\eqref{eq:ratio_BSM}. In fact, this is not the case for the lifetime ratio $\tau(B_s)/\tau(B_d)$, where SU(3)$_F$ breaking effects play a dominant role. Therefore, 
while the 
presence of NP in $b \to c \bar u d (s)$ transitions would also affect the theoretical prediction of $\tau(B_s)/\tau(B_d)$, 
a detailed study of BSM contributions to this ratio would currently be strongly limited by the size of the non-perturbative input which parametrise the matrix elements of two-quark operators, and even more of the corresponding SU(3)$_F$ breaking effects, which are poorly known, particularly for the Darwin operator, see e.g.\ the recent work~\cite{Lenz:2022rbq}. Hence, given the current status of the SM prediction, we postpone the study of NP effects in $\tau(B_s)/\tau(B_d)$ to a future work, once further insights on the size of matrix element of the Darwin operator and 
of the SU(3)$_F$ breaking effects will become available. 
\\
The decay rate differences $\Delta \Gamma_q$ and the semileptonic CP asymmetries $a_{sl}^q$, with $q = d,s,$
are related respectively to the real and the imaginary part of the ratio of the absorptive and dispersive contributions to the mixing of the neutral $B_q$ mesons, see e.g.\ Refs.~\cite{Buras:1984pq,Proceedings:2001rdi}
 \begin{eqnarray}
 \frac{\Delta \Gamma_q}{\Delta M_q}
 =  - {\rm Re}
 \left(
 \frac{\Gamma_{12}^q}{M_{12}^q} 
 \right)
 \, ,  & \quad &
 a_{sl}^q  =  {\rm Im}
 \left(
 \frac{\Gamma_{12}^q}{M_{12}^q} 
 \right)
 \,,
 \label{eq:asl-q}
 \end{eqnarray}
 where it is convenient to rewrite $\Gamma_{12}^q /M_{12}^q $
 as \cite{Beneke:2003az,Lenz:2006hd,Artuso:2015swg}
 \begin{equation}
 - \frac{\Gamma_{12}^q}{M_{12}^q}
 = 
 \frac{\Gamma_{12}^{q,cc}}{\tilde{M}_{12}^q}
 + 2 \frac{\lambda_u^q}{\lambda_t^q}
 \frac{\Gamma_{12}^{q,cc}-\Gamma_{12}^{q,cu} }{\tilde{M}_{12}^q}
 + \left( \frac{\lambda_u^q}{\lambda_t^q} \right)^2
 \frac{\Gamma_{12}^{q,cc}-2 \Gamma_{12}^{q,cu} + \Gamma_{12}^{q,uu}}{\tilde{M}_{12}^q}\,.
 \label{eq:G12M12}
  \end{equation}
In Eq.~\eqref{eq:G12M12}, $\lambda_x^q = V_{xq}^* V_{xb}$ and we have introduced the notation
$M_{12}^q = (\lambda_t^q)^2 \tilde{M}_{12}^q$. Moreover, $\Gamma_{12}^{q,xy}$ corresponds to the imaginary part of the mixing diagram with internal quarks $x$ and~$y$, cf.\ Figure~\ref{fig:mixing-NP}. From Eq.~\eqref{eq:asl-q} and 
taking into account both the CKM- and the GIM-suppression~\cite{Glashow:1970gm}, it becomes apparent that while the decay rate difference is dominated by the first term on the r.h.s.\ of Eq.~\eqref{eq:G12M12}, where no suppression is present, 
the prediction for~$a_{sl}^q$ is given by the second and third terms on the r.h.s.\ of Eq.~\eqref{eq:G12M12}, which are instead strongly suppressed. 
However, under the assumption of NP in $b \to c \bar u d(s)$ transitions only, the GIM suppression would be clearly lifted, leading to large effects in the corresponding theoretical predictions for $a_{sl}^q$, but only a small correction to the SM value of $\Delta \Gamma_q$. Hence, in this specific BSM scenario, we expect our bounds from mixing observables to be dominantly driven by $a_{sl}^q$, while we 
do not expect any significant constraint from $\Delta \Gamma_q$.
\\
Based on the calculations in Refs.~\cite{Davies:2019gnp,Dowdall:2019bea,FermilabLattice:2016ipl,DiLuzio:2019jyq,King:2019lal,Kirk:2017juj,Grozin:2016uqy,Lenz:2006hd,Beneke:2003az,Beneke:1998sy,Beneke:1996gn,Dighe:2001gc}, the most recent SM predictions for $a_{sl}^q$ and $\Delta \Gamma_q$, respectively, read~\cite{Lenz:2019lvd}~\footnote{See also Ref.~\cite{Gerlach:2022hoj}
for new results for $\Delta \Gamma_s$ including NNLO-QCD corrections.}
 \begin{eqnarray}
 a_{sl}^{d, {\rm HQE}} = (-4.73 \pm 0.42 ) 
 \cdot 10^{-4} 
 \, ,
 & \quad  &
 a_{sl}^{s, {\rm HQE}} = (2.06 \pm 0.18 ) 
 \cdot 10^{-5} 
 \, ,
 \end{eqnarray}
 \begin{eqnarray}
 \Delta \Gamma_d^{\rm HQE}  =  (2.6 \pm 0.4) 
 \cdot 10^{-3} \, \mbox{ps}^{-1}\, ,
 & \quad & 
 \Delta \Gamma_s^{\rm HQE}  = (9.1 \pm 1.3)
 \cdot 10^{-2} \, \mbox{ps}^{-1}
 \,.
\end{eqnarray}
 Unfortunately, the semileptonic CP asymmetries have not been measured yet and HFLAV \cite{Amhis:2022mac}
 quotes for the current experimental bounds
 \begin{eqnarray}
 a_{sl}^{d,{\rm Exp.}} = (-21 \pm 17 ) 
 \cdot 10^{-4} 
 \, ,
 &\quad&
 a_{sl}^{s, {\rm Exp.}} = (-60 \pm 280 ) 
 \cdot 10^{-5} 
 \, ,
 \label{eq:asl_HFLAV}
 \end{eqnarray}
and similarly for $\Delta \Gamma_d$, where HFLAV  presents again only a bound, while  $\Delta \Gamma_s$ is 
measured quite precisely \cite{Amhis:2022mac}
\begin{eqnarray}
 \left(\frac{\Delta \Gamma_d}{\Gamma_d}\right)^{\rm Exp.}  =  0.001 \pm 0.010 \, ,
 &\quad&
 \Delta \Gamma_s^{\rm Exp.}  = (8.4 \pm 0.5)
 \cdot 10^{-2} \, \mbox{ps}^{-1}
 \, .
 \label{eq:DGd}
\end{eqnarray}
Despite the current experimental ignorance about the precise values of the semileptonic CP asymmetries, the quoted loose bound on $a_{sl}^d$ poses already some interesting limits on the size of potential BSM effects.
Obviously any experimental improvement on the values in
Eq.~\eqref{eq:asl_HFLAV} would have a severe impact on the results of our
study.
At the LHC a future precision of 
$\pm \, 3 \cdot 10^{-4}$ for the semileptonic CP
asymmetries is foreseen for the high luminosity
upgrade (Upgrade II) \cite{Cerri:2018ypt}, while a future
FCC-ee collider running at the $Z$ resonance could
even go one order of magnitude further and reach a
precision of $\pm \, 3 \cdot 10^{-5}$~\cite{Monteil}. Moreover, LHCb expects to reduce the uncertainty of $\Delta \Gamma_d / \Gamma_d$ in Eq.~(\ref{eq:DGd}) by one order of magnitude, namely to $\pm \, 0.001$~\cite{LHCb:2018roe}.
\\
To specify the impact that future experimental improvements would have on our constraints for the NP Wilson coefficients
and to identify the key observables,
in our numerical analysis we use both the current experimental bounds for the mixing observables, as well as the future projections.
Specifically, we consider 
\begin{eqnarray}
 a_{sl}^{d, \rm Proj. \, A} = (-21 \pm 3 ) 
 \cdot 10^{-4} 
 \, ,
 & \quad &
 a_{sl}^{d, \rm Proj. \, B} = (-21.0 \pm 0.3 ) 
 \cdot 10^{-4} 
 \, ,
 \label{eq:asl_HFLAV-proj}
 \\[2mm]
 \left(\frac{\Delta \Gamma_d}{\Gamma_d}\right)^{ \rm Proj.} & = &
 0.001 \pm 0.001
 \, .
 \label{eq:DeltaGamma_HFLAV-proj}
 \end{eqnarray}
 The main result of this paper 
 is the computation of 
 the analytic expressions for the leading contributions of
 the generalised effective BSM Hamiltonian to the
 lifetime
 and to the mixing observables,
 cf.\ Figures~\ref{fig:PI-WE-NP} and  \ref{fig:mixing-NP}, in the presence of NP in 
 $b \to c \bar{u} d (s)$, namely for the case of
 one massive internal particle in the corresponding diagrams.
 Hence,
 our work complements earlier results obtained in
 Refs.~\cite{Jager:2019bgk, Jager:2017gal} for
 two equal massive internal particles. The case of
 two massless internal particles can be
 trivially obtained from our results as well as from those in Refs.~\cite{Jager:2019bgk, Jager:2017gal}, by
 taking the limit $m_c \to 0$. As a further
 consistency check we have in addition
 re-calculated all the expressions for two equal massive
 internal particles obtained in
 Refs.~\cite{Jager:2019bgk, Jager:2017gal}. 
 We have found exact agreement, except for some
 mixing contributions that were overlooked in
 Ref.~\cite{Jager:2019bgk} and therefore we present the missing
 contributions in our paper.
 To illustrate the phenomenological impact of
 our results we investigate the scenario in which
 NP is acting only in the decay channels 
 $b \to c \bar{u} d (s)$. A detailed 
 BSM analysis of the more general case in which NP is acting in different decay channels -- including a connection to the {$B$ anomalies} via BSM effects in $b \to c \bar{c} s$ transitions, as well as the study of UV complete models that could lead to these effects -- is postponed to a future work.
\\
The paper is organised as follows: in Section \ref{NP_in_cud} we introduce BSM effects in tree-level non-leptonic $b \to c \bar u d (s)$ decays. Specifically, we describe the general effective Hamiltonian 
in Section \ref{Heff}, in Section
\ref{non-leptonic} we briefly review the results of Ref.~\cite{Cai:2021mlt}, where this effective Hamiltonian has been applied to the study of non-leptonic $B$-meson decays like $\bar B_s \to D_s^+ \pi^-$, while
our analytic expressions for the BSM contributions to the lifetime ratio $\tud$
and to the mixing observables $\Delta \Gamma_q$ and $a_{sl}^q$ are presented in Sections~\ref{Lifetime} and~\ref{asl}, respectively.
The numerical study of BSM effects, in the presence of NP in the decay channels $b \to c \bar u d (s)$ only, is described in Section~\ref{Numerics}, where we compare
our bounds on the NP Wilson coefficients obtained from the lifetime ratio $\tud$ and the semileptonic CP asymmetry $a_{sl}^d$, with those from the
non-leptonic $B$-meson decays shown in Ref.~\cite{Cai:2021mlt}. 
Finally, in
Section~\ref{Conclusion} we summarise our results and give an outlook on future developments.

\section{\boldmath  NP in $b \to c \bar u d (s)$ transitions}
\label{NP_in_cud}
\subsection{General effective Hamiltonian}
\label{Heff}
In our analysis we assume that the SM weak effective Hamiltonian describing the tree-level non-leptonic transition $b \to c \bar u d$ \footnote{For the sake of a cleaner notation, we explicitly consider only the mode $b \to c \bar u d$. In fact, 
everything holds in the same way for $b \to c \bar u s$, given that one replaces in all expressions $d \to s$.}, at the scale $\mu \sim m_b$, is extended  with the following model-independent effective Hamiltonian ${\cal H}^{\rm NP}_{\rm eff}$,
see e.g. Refs.~\cite{Cai:2021mlt, Jager:2019bgk}
\begin{equation}
{\cal H}_{\rm eff}^{\rm NP}(x)
= 
\frac{4 G_F}{\sqrt 2} V_{c b} V_{u d}^*
\sum_{i = 1}^{10} 
\left[
C_i^{\rm NP} \, Q_i (x) + C_i^{\prime {\rm NP}} \, Q_i^\prime (x) 
\right] 
+ {\rm h.c.},
\label{eq:Heff-NP}    
\end{equation}
where the NP four-quark operators $Q_i$ are defined as
\footnote{Note that our notation for the NP operator basis is different, both in the order and in the labeling, from others used in the literature, see e.g.\ Refs.~\cite{Cai:2021mlt, Jager:2019bgk}.  
We denote with odd numbers colour singlets and with even number colour-rearranged operators, moreover
we start with the SM Dirac structure, and then add further (axial-)vector, (pseudo-)scalar and finally tensor structures.
}
\begin{align}
& Q_1 = (\bar c^i \gamma_\mu P_L \, b^i) (\bar d^j \gamma^\mu P_L \, u^j)\,,
\quad &
& Q_2 = (\bar c^i \gamma_\mu P_L \, b^j) (\bar d^j \gamma^\mu P_L \, u^i)\,,
\label{eq:Q1-Q2}
\\[2mm]
& Q_3 = (\bar c^i \gamma_\mu P_R \, b^i) (\bar d^j \gamma^\mu P_L \, u^j)\,,
\quad &
& Q_4 = (\bar c^i \gamma_\mu P_R \, b^j) (\bar d^j \gamma^\mu P_L \, u^i)\,,
\label{eq:Q3-Q4}
\\[2mm]
& Q_5 = (\bar c^i P_L \, b^i) (\bar d^j P_R \, u^j)\,,
\quad &
& Q_6 = (\bar c^i P_L \, b^j) (\bar d^j P_R \, u^i)\,,
\label{eq:Q5-Q6}
\\[2mm]
& Q_7 = (\bar c^i P_R \, b^i) (\bar d^j P_R \, u^j)\,, 
\quad &
& Q_8 = (\bar c^i P_R \, b^j) (\bar d^j P_R \, u^i)\,,
\label{eq:Q7-Q8}
\\[2mm]
& Q_9 = (\bar c^i \sigma_{\mu\nu} P_R\, b^i)(\bar d^j \sigma^{\mu\nu} P_R\, u^j)\,,
\quad &
& Q_{10} = (\bar c^i \sigma_{\mu\nu} P_R\, b^j)(\bar d^j \sigma^{\mu\nu} P_R\, u^i)\,.
\label{eq:Q9-Q10}
\end{align}
Here $P_{L,R} = (1 \mp \gamma_5)/2$, $\sigma^{\mu \nu} = (i/2)[\gamma^\mu, \gamma^\nu]$, and $i, j = 1, 2, 3$ label the $SU(3)_c$ indices. The remaining operators $Q_i^\prime$ in Eq.~\eqref{eq:Heff-NP} are obtained from those in Eqs.~\eqref{eq:Q1-Q2} - \eqref{eq:Q9-Q10} by exchanging the corresponding
chirality, i.e. $Q_i^\prime = Q_i|_{P_L \leftrightarrow P_R} $.
Note that $Q_1$ and $Q_2$ in Eq.~\eqref{eq:Q1-Q2} coincide with the SM operators, however their Wilson coefficients in Eq.~\eqref{eq:Heff-NP} originate only from NP effects. In fact, the SM weak effective Hamiltonian reads~\cite{Buchalla:1995vs}
\begin{equation}
{\cal H}_{\rm eff}^{\rm SM}(x) = \frac{4 G_F}{\sqrt 2} V_{c b} V_{ud}^* \left[ C_1^{\rm SM} Q_1(x) +  C_2^{\rm SM} Q_2(x)\right] + {\rm h.c.}\,,
\label{eq:Heff-SM}    
\end{equation}
and for brevity we introduce the notation
\begin{equation}
    {\cal H}_{\rm eff} \equiv {\cal H}_{\rm eff}^{\rm SM} + {\cal H}_{\rm eff}^{\rm NP}\,, 
    \qquad 
    \vec C \equiv \vec C^{ \rm SM} + \vec C^{ \rm NP}\,, 
    \qquad 
    \vec C^{\prime} \equiv \vec C^{{\prime}\, \rm NP}\,.
    \label{eq:Heff-SM-and-NP}
\end{equation}
The scale dependence of the Wilson coefficients $C^{(\prime)}_i$ is governed
by the renormalisation-group equations (RGEs)
\begin{equation}
\mu \frac{d}{d \mu} C_i^{(\prime)} (\mu) = (\gamma^T)_{ij} (\mu) \, C_j^{(\prime)} (\mu),
\qquad i, j = 1, \ldots, 10\,,
\label{eq:RGE}
\end{equation}
where $ \hat \gamma (\mu)$ denotes the $10 \times 10$ anomalous dimension matrix
(ADM) \cite{Ciuchini:1997bw, Ciuchini:1998ix, Buras:2000if}. Note that due to parity conservation in QCD, primed and unprimed Wilson coefficients follow the same RGEs and do not mix with each other under renormalisation.
The solution of Eq.~\eqref{eq:RGE} can be presented as
\begin{equation}
C^{(\prime)}_i (\mu) = U_{ij} (\mu, \mu_0) \, C^{(\prime)}_j (\mu_0)\,,
\label{eq:RGE-solution}
\end{equation}
where $\hat U (\mu, \mu_0)$ is the evolution matrix describing 
the running of the Wilson coefficients from the scale 
$\mu_0$ to the scale $\mu$.
 The explicit expression for $\hat U (\mu, \mu_0)$ can be found e.g.\ in Ref.~\cite{Jager:2019bgk}.

\begin{figure}[t]
\centering
   \begin{subfigure}{0.65\textwidth}
   \includegraphics[scale=0.45]{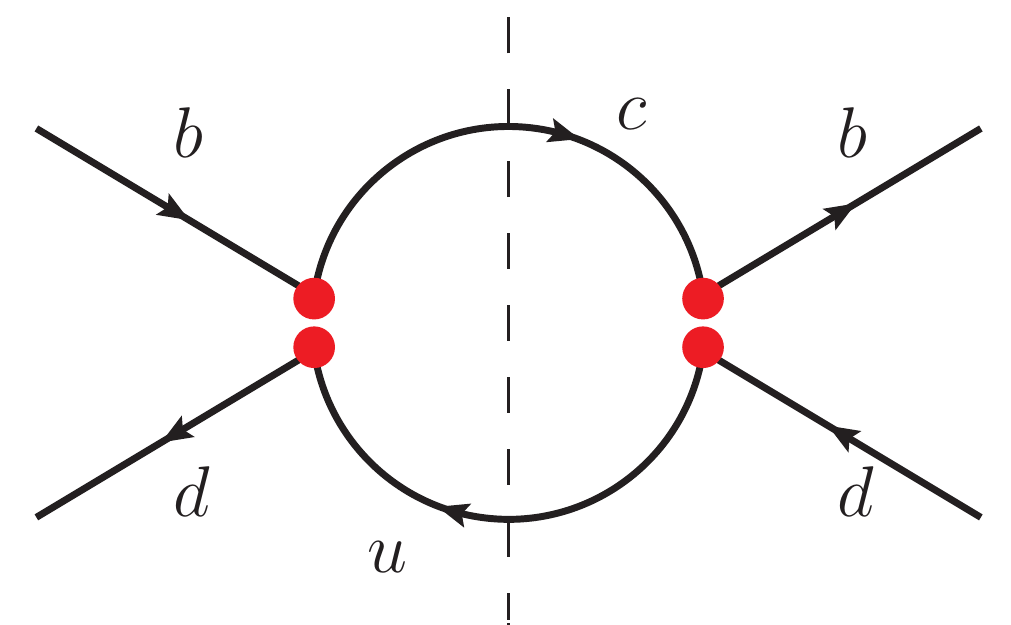}
   \qquad 
   \includegraphics[scale=0.45]{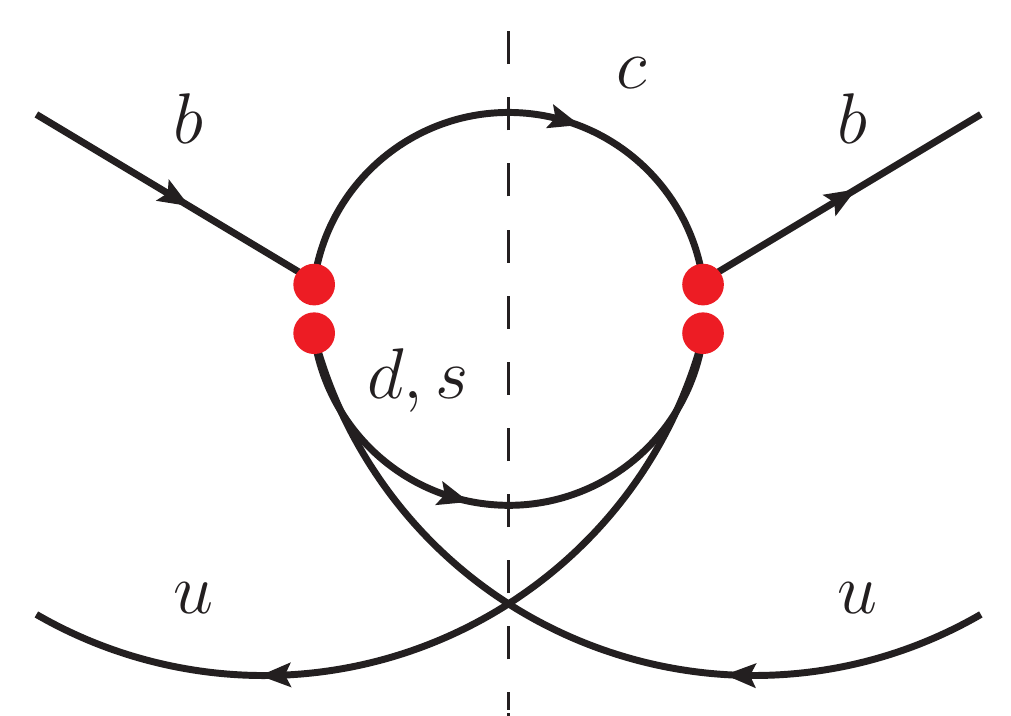}
   \caption{}
   \label{fig:PI-WE-NP}
   \end{subfigure}
   \begin{subfigure}{0.30\textwidth}
   \includegraphics[scale=0.45]{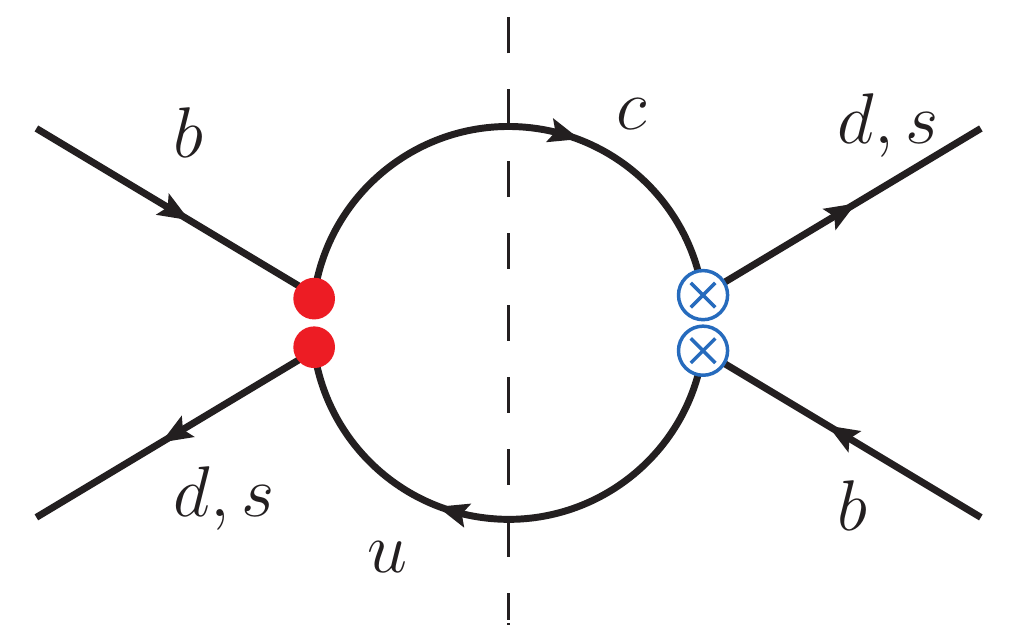}
   \caption{}
   \label{fig:mixing-NP}
   \end{subfigure}
   \caption{(a) Leading diagrams contributing to the total width difference of the $B_d$ (left) and of the $B^+$ (right) mesons.
   (b) Leading diagram contributing to $\Gamma_{12}^{q, cu}$.
    The double red circles indicate the insertion of one of the 20 operators in effective Hamiltonian in Eq.~\eqref{eq:Heff-SM-and-NP}, while
    the double crossed blue circles denote one of
    the SM operators from the Hamiltonian in Eq.~\eqref{eq:Heff-SM}.
    In fact, the mixing diagram (b) originates from the interference of the $b \to c \bar{u} d(s)$
    and $b \to u \bar{c} d(s)$ transitions, and under the assumption of NP effects only in $b \to c \bar{u} d (s)$ decays, at least one of the two operators inserted always corresponds to the SM one $Q_{1,2}$.
    }
\end{figure}

\subsection{BSM effects in non-leptonic \textit{B}-meson decays} 
\label{non-leptonic}
Starting from the general NP effective Hamiltonian, cf.~Eq.~(\ref{eq:Heff-NP}), the authors of Ref.~\cite{Cai:2021mlt} have investigated the non-leptonic $B$-meson decays
$\bar{B}_{(s)}^0 \to {D}_{(s)}^{(*)+} L^-$, with $L \in \{ \pi, \rho, K^{(*)}\}$,  
within the QCDF framework and by also computing NLO-QCD corrections.
They have found that new contributions to the operators $Q_{1,2}$ 
could explain the observed deviations of the QCDF prediction from the corresponding experimental data at the $1 \sigma$ level, whereas new
contributions to the operators
$Q_{7,8}$ and $Q_{5,6}^\prime$ could accommodate the data within $2\sigma$. In order to facilitate the comparison, in Section~\ref{Numerics}, we overlay our constraints for the favoured regions of the NP Wilson coefficients originating from the $B$-meson lifetimes and $B$-mixing, with the ones obtained in Ref.~\cite{Cai:2021mlt}.

\subsection{\boldmath BSM effects to lifetimes}
\label{Lifetime}
By means of the optical theorem, the total decay width of a $B$ meson can be related to the imaginary part of the forward-scattering matrix element of the time-ordered product of the double insertion of the effective Hamiltonian, i.e.
\begin{equation}
    \Gamma (B) = \frac{1}{2 m_B}{\rm Im} \langle B|  {\cal T} | B \rangle \,,
    \label{eq:GammaB}
\end{equation}
with the transition operator given by
\begin{equation}
    {\cal T} = i \int d^4 x \, T\, \{ {\cal H}_{\rm eff}(x), {\cal H}_{\rm eff}(0)\}\,.
    \label{eq:T-operator}
\end{equation}
Within the HQE, the non-local operator in Eq.~\eqref{eq:T-operator} can be expressed as a systematic expansion in inverse powers of the heavy $b$-quark mass,
leading to the following series 
\begin{equation}
\Gamma(B) = 
\Gamma_3  +
\Gamma_5 \frac{\langle {\cal O}_5 \rangle}{m_b^2} + 
\Gamma_6 \frac{\langle {\cal O}_6 \rangle}{m_b^3} + \ldots  
 + 16 \pi^2 
\left( 
  \tilde{\Gamma}_6 \frac{\langle \tilde{\mathcal{O}}_6 \rangle}{m_b^3} 
+ \tilde{\Gamma}_7 \frac{\langle \tilde{\mathcal{O}}_7 \rangle}{m_b^4} + \ldots
\right)\,,
\label{eq:HQE}
\end{equation}
where $\Gamma_d$ are short-distance functions which can be computed perturbatively in QCD,
and $\langle {\cal O}_d \rangle \equiv
\langle B | {\cal O}_d |B  \rangle/(2 m_{B})$ denote the matrix element of local $\Delta B = 0$ operators of increasing dimension $d$. We emphasise that starting from dimension-six, both two- and four-quark operator contributions appear, with the latter originating from loop-enhanced diagrams, as reflected by the explicit factor of $16 \pi^2$ in Eq.~\eqref{eq:HQE}. More details on the structure of the HQE for the $b$-system, as well as a complete list of references can be found e.g.\ in Ref.~\cite{Lenz:2022rbq}.
Here, it is sufficient to stress that the lifetime ratio $\tau(B^+)/\tau(B_d)$ is only sensitive to the effect of four-quark operators since, in the limit of isospin symmetry, the contribution of two-quark operators cancels exactly from the r.h.s.\ of Eq.~\eqref{eq:ratio_BSM}. It thus follows that, in the presence of NP parametrised by the effective Hamiltonian in Eq.~\eqref{eq:Heff-NP}, 
 the dominant correction to the lifetime ratio $\tud$ arises at order $1/m_b^3$ in the HQE, from the discontinuity of the one-loop diagrams shown in Figure~\ref{fig:PI-WE-NP} corresponding to the weak exchange (WE) and Pauli interference (PI) topologies, respectively. 
Specifically, at dimension-six and at LO-QCD, the contribution of the NP operators in Eq.~\eqref{eq:Heff-NP} 
to the imaginary part of the transition operator in Eq.~\eqref{eq:T-operator} can be compactly written as \footnote{The NP Wilson coefficients $C_i^{\rm NP}$ can in general be complex numbers, however, in our numerical analysis we assume them to be real i.e.\ $C_i= C_i^*$.}
\begin{equation}
{\rm Im}{\cal T}_{\rm NP}^{X, c q} = \frac{G_F^2 m_b^2}{6 \pi}|V_{ud}|^2|V_{c b}|^2 \, (1 - \rho)^2
\left[\,
\sum_{m, n = 1}^{20} C_m \, C_n^* \, A^{X, c q}_{m n}
- \sum_{m,n = 1}^{2}
C_m^{\rm SM} C_n^{\rm SM} \, A^{X,c q}_{mn}\,
\right].
\label{eq:T-NP}
\end{equation}
Here $X =$ PI, WE, while $c, q,$ label the internal quarks running into the loop, with $q = u, d$. As we neglect the mass of the light quarks $m_{u,d,s} = 0$, there is only one dimensionless mass parameter $\rho = m_c^2/m_b^2$. Moreover, in Eq.~\eqref{eq:T-NP}, for brevity, we have introduced the notation
\begin{equation}
     C^{\rm NP}_m \equiv C^{^\prime \rm NP}_{m-10}, \quad m > 10\,.
\end{equation}
The functions~$A^{X, cq}_{mn}$ in Eq.~\eqref{eq:T-NP} represent linear combinations, with coefficients depending only on the parameter
$\rho$, of the dimension-six $\Delta B = 0$ four-quark operators 
$\Oq_i$ and $\Oqp_i$. The latter are defined, respectively, as
\begin{align}
  \Oq_1
  & =  
  4\, (\bar b \, \gamma_\mu P_L \, q)\,(\bar{q} \, \gamma^\mu P_L \, b) ,
  &
  \Oq_2  
  & =  
  4 \, (\bar b P_L \, q)\,(\bar{q} P_R \, b) ,
  \label{eq:O1-O2-HQET} 
  \\[1mm]
  \Oq_3  
  & =  
  4 \,(\bar b \, \gamma_\mu P_L \, t^a q)\,(\bar{q}\,\gamma^\mu P_L\,t^a b),
  &
  \Oq_4 
  & =  
  4\,(\bar b P_L \, t^a q)\,(\bar{q} P_R \, t^a b),
  \label{eq:O3-O4-HQET} 
  \\[1mm]
  \Oq_5  
  & =  
  4\,(\bar b \, \gamma_\mu P_L \, q)\, (\bar{q} \, \gamma^\mu P_R \, b) ,
  &
  \Oq_6  
  & =  
  4 \, (\bar b P_L \, q) \, (\bar{q} P_L \, b) ,
  \label{eq:O5-O6-HQET} 
  \\[1mm]
  \Oq_7  
  & =  
  4\,(\bar b \, \gamma_\mu P_L \, t^a q) (\bar{q}\,\gamma^\mu P_R\,t^a b),
  &
  \Oq_8 
  & =  
  4 \, (\bar b P_L \, t^a q) \,
  (\bar{q}P_L \, t^a b)\,,
  \label{eq:O7-O8-HQET}
\end{align}
while the primed operators $\Oqp_i$ are obtained by replacing $P_L \leftrightarrow P_R$ in Eqs.~\eqref{eq:O1-O2-HQET} - \eqref{eq:O7-O8-HQET}. 
In the above basis, $q$ indicates the external light quark, i.e.\ $q = u,d$. However, for the sake of a simpler notation, we omit to include a label on the corresponding operator. In fact, the quark content can be easily understood, since at the order we are considering, only four-quark operators with external $u$-quark contribute to the total width of the $B^+$ meson, whereas those with external $d$-quark to that of the $B_d$ \footnote{In fact, the contribution of four-quark operators, in which the external light quark differs from the spectator quark in the corresponding $B$ meson, constitutes a subleading effect which goes beyond the current accuracy of our analysis, see e.g. Ref.~\cite{Lenz:2022rbq}.}, and note that for the same reason, the $b \to c \bar u s$ transition can contribute only to $\Gamma (B^+)$. In addition, due to isospin symmetry, the matrix element of operators with external $u$- or $d$-quark are parametrised in terms of the same set of Bag parameters, cf.\ Eq.~\eqref{eq:Par-ME}.
In Eqs.~\eqref{eq:O1-O2-HQET} - \eqref{eq:O7-O8-HQET}, a summation over colour indices is understood while in Eqs.~\eqref{eq:O3-O4-HQET}, \eqref{eq:O7-O8-HQET}, $t^a_{ij}$, with $a = 1, \ldots, 8$, denote the $SU(3)_c$ generators. The latter satisfy the following completeness relation
\begin{equation}
    t_{ij}^a t^a_{lk} = \frac12 \left( \delta_{ik}\delta_{jl} - \frac{1}{N_c} \delta_{ij}\delta_{lk} \right)\,,
\end{equation}
which can be used to express the $\Delta B = 0$ colour-rearranged operators in terms of the corresponding colour-singlet and colour-octet ones. We emphasise that the basis of $\Delta B = 0$ operators in Eqs.~\eqref{eq:O1-O2-HQET} - \eqref{eq:O7-O8-HQET} does not include tensor operators as these can be rewritten, by using the equations of motion for the $b$-quark spinor and up to power corrections, as a linear combination of operators containing the remaining Dirac structures, e.g. 
\begin{align}
(\bar b \sigma^{\rho \tau} q)(\bar q \sigma_{\rho \tau} b) = 
 - 2 \left[(\bar b  q)(\bar q  b) 
 - (\bar b \gamma^\rho q) (\bar q \gamma_\rho b) 
 + (\bar b \gamma_5 q)(\bar q \gamma_5 b)  
 + (\bar b \gamma^\rho \gamma_5 q)(\bar q \gamma_\rho \gamma_5 b)\right] + {\cal O}\left(\frac{1}{m_b}\right)\,.
\nonumber\\
\end{align}
For $\Gamma(B^+)$, the dominant contribution comes from the PI topology, cf. r.h.s.\ of Figure~\ref{fig:PI-WE-NP}. In this case the analytic expressions for the entries $A^{{\rm PI},cd}_{mn}$ in Eq.~\eqref{eq:T-NP} respectively read
\begin{eqnarray}
& \displaystyle
A_{1,2}^{{\rm PI}, cd}  = 3 \, \Oq_1, \qquad  
A_{1,4}^{{\rm PI}, cd} = - \frac{3}{2} \sqrt \rho \, \Oq_1\,,  \qquad
A_{1,6}^{{\rm PI}, cd}   = - \frac{3}{4} \sqrt \rho \, \left[\Oq_5 - 2\,\Oq_6\right]\,, 
&
\nonumber \\
& \displaystyle
A_{1,8}^{{\rm PI}, cd} = - \frac{3}{4} \left[\Oq_5 - 2 \, \Oq_6 \right], \qquad
A_{1,10}^{{\rm PI}, cd}  = 9 \left[\Oq_5 - 2 \, \Oq_6 \right]\,, 
&
\end{eqnarray}
\begin{eqnarray}
& \displaystyle
A_{3,4}^{{\rm PI}, cd}  = \frac{1}{2} \left[(1 + 2 \rho) \, \Oq_1 
- 2 (2 + \rho) \, \Oq_2 \right]\,, \qquad
A_{3,6}^{{\rm PI}, cd} = - \frac{1}{4} \left[(1 - \rho) \Oq_5 + 2 (1 + 2 \rho) \Oq_6 \right] \,,
&
\nonumber \\
& 
\displaystyle
A_{3,8}^{{\rm PI}, cd} = -\frac{3}{2} \sqrt \rho \, \Oq_6\,, \qquad 
A_{3,10}^{{\rm PI}, cd}  = - 6 \sqrt \rho \, \left[\Oq_5 - \Oq_6\right]\,,
&
\end{eqnarray}
\begin{eqnarray}
& \displaystyle
A_{5,6}^{{\rm PI}, cd}  = \frac{1}{8} \left[(1 + 2 \rho) \Oqp_1 - 2 (2 + \rho) \Oqp_2 \right]\,, \qquad
A_{5,8}^{{\rm PI}, cd} = \frac{3}{8} \sqrt \rho \left[\Oqp_1 - 2 \Oqp_2 \right]\,, 
&
\nonumber \\
& \displaystyle
A_{5,10}^{{\rm PI}, cd} = - \frac{3}{2} \sqrt \rho \left[\Oqp_1 + 2 \Oqp_2 \right]\,,
&
\end{eqnarray}
\begin{eqnarray}
& \displaystyle
A_{7,8}^{{\rm PI}, cd}  = 
\frac{1}{8} \left[(2 + \rho) \Oqp_1 - 2 \, (1 + 2 \rho) \Oqp_2 \right]\,,
\qquad 
A_{7, 10}^{{\rm PI}, cd} = - \frac{1}{2} \left[(4 - \rho) \Oqp_1 
+ 2 \, (1 + 2 \rho) \Oqp_2 \right]\,, \qquad
&
\end{eqnarray}
\begin{eqnarray}
& \displaystyle
A_{9,10}^{{\rm PI}, cd}  = 
2 \left[ (14 + \rho) \Oqp_1 - 2 \, (1 + 2 \rho) \Oqp_2 \right]\,,
&
\end{eqnarray}
while the remaining coefficient functions are obtained as follows
\begin{equation}
\begin{tabular}{lr}
$ \ast \, A^{{\rm PI}, cd}_{m, \, n} = 0$\,,
& 
$
\left\{
\begin{array}{l}
     m = 1, \ldots, 10, \,\, n = 11, \ldots 20\,, \\
     m = 11, \ldots, 20, \,\, n = 1, \ldots 10\,,
\end{array}
\right.
$
\\[5mm]
$ \ast \, A^{{\rm PI}, cd}_{2m-1, \, 2n-1} = A^{{\rm PI}, cd}_{2m, \,2n} = 
 A^{{\rm PI}, cd}_{2m-1, \, 2n} 
\Big|_{\Oq^{(\prime)}_i \to \left(\frac{\Oq^{(\prime)}_i}{N_c} + 2 \, \Oq^{(\prime)}_{i+2}\right)}\,, $
& $m, n = 1, \ldots, 5\,,$  
\\
$\ast \, A^{{\rm PI}, cd}_{2m, 2n-1} = A^{{\rm PI}, cd}_{2m- 1, 2n}\,,$ 
& 
$m,n = 1, \ldots, 5\,,$
\\[4mm]
$\ast \, A^{{\rm PI}, cd}_{m, n} = A^{{\rm PI}, cd}_{n, m}\,,$ 
& 
$
\left\{
\begin{array}{l}
     m, n = 1, \ldots, 4\,,\\
     m, n  = 5, \ldots, 10\,,
\end{array}
\right.
$
\\[7mm]
$\ast \, A^{{\rm PI}, cd}_{m, n} = A^{{\rm PI}, cd}_{n, m}\Big|_{\Oq_i \leftrightarrow \Oqp_i}\,,$ 
& 
$
\left\{
\begin{array}{l}
     m = 1, \ldots, 4, \,\, n = 5,  \ldots, 10 \,, \\
     m = 5, \ldots, 10, \,\, n = 1, \ldots, 4 \,, 
\end{array}
\right.
$
\\[6mm]
$\ast \, A^{{\rm PI}, cd}_{m,n} =A^{{\rm PI}, cd}_{m-10, \, n-10} 
\Big|_{\Oq_i \leftrightarrow \Oqp_i}\,,$
& 
$m,n = 11, \, \ldots, 20 \,.$ 
\end{tabular}
\end{equation}
It is worth emphasising that the corresponding contribution of the decay channel $b \to c \bar{u}s$ to  Eq.~\eqref{eq:T-NP}, with $q = s$, is straightforwardly obtained by setting 
 $A^{{\rm PI}, cs}_{m,n} =  
 A^{{\rm PI}, cd}_{m,n}, $ and by replacing the proper CKM factor, i.e.\ $|V_{ud}|^2 \to |V_{us}|^2$.
\\
In the case of $\Gamma(B_d)$, the dominant contribution comes from the WE topology, cf.\ l.h.s.\ of Figure~\ref{fig:PI-WE-NP}, and the corresponding analytic expressions for $A_{mn}^{{\rm WE}, cu }$ in Eq.~\eqref{eq:T-NP} read
\begin{eqnarray}
\label{eq:WE-1}
& \displaystyle
A_{1,2}^{{\rm WE},cu} = -\frac{1}{2} (2 + \rho) \, \Oq_1 + (1 + 2 \rho) \Oq_2\,, \qquad 
A_{1,4}^{{\rm WE},cu} = - 3 \sqrt \rho \, \Oq_2\,, 
&
\nonumber \\[1.5mm]
& \displaystyle
A_{1,16}^{{\rm WE},cu}  = \frac{1}{4}
\left[(2 + \rho) \, \Oqp_5 - 2 \, (1 + 2 \rho) \Oqp_6 \right]\,, 
\qquad 
A_{1,18}^{{\rm WE},cu} = \frac{3}{4} \sqrt \rho \, [\Oqp_5 - 2 \Oqp_6]\,,
&
\\[1.5mm]
& \displaystyle
A_{1,20}^{{\rm WE},cu} = - 3 \sqrt \rho \, [\Oqp_5 + 2 \Oqp_6]\,,
& \nonumber
\end{eqnarray}
\begin{eqnarray}
& \displaystyle
A_{3,4}^{{\rm WE},cu} = 6 \, \Oq_2\,,  \qquad 
A_{3,16}^{{\rm WE},cu} = \frac{3}{2} \sqrt \rho \, \Oqp_6\,, \qquad 
A_{3,18}^{{\rm WE},cu} = \frac 3 2 \, \Oqp_6\,, \qquad 
A_{3,20}^{{\rm WE},cu} = 18 \,  \Oqp_6\,, \quad 
&
\end{eqnarray}
\begin{eqnarray}
& \displaystyle
A_{5,6}^{{\rm WE},cu} = - \frac{1}{8} \left[(2 + \rho) \, \Oq_1 - 2 \, (1 + 2 \rho) \Oq_2\right]\,, \qquad
A_{5,8}^{{\rm WE},cu} = - \frac 3 8 \sqrt \rho \, [\Oq_1 - 2 \, \Oq_2]\,, 
& \nonumber
\\[1.5mm]
& \displaystyle
A_{5,10}^{{\rm WE},cu} = \frac 3 2 \sqrt \rho \, [\Oq_1 + 2 \, \Oq_2]\,,  
& 
\end{eqnarray}
\begin{eqnarray}
& \displaystyle
A_{7,8}^{{\rm WE},cu} = \frac{1}{8} \left[2 \, (2 + \rho) \, \Oq_2 
- \, (1 + 2 \rho) \Oq_1 \right]\,, \quad
A_{7,10}^{{\rm WE},cu} = \frac{1}{2} \left[(1 + 2\, \rho) \, \Oq_1 
+ 2 \, (4 - \rho) \, \Oq_2 \right]\,,
\nonumber\\
\end{eqnarray}
\begin{eqnarray}
& \displaystyle
A_{9,10}^{{\rm WE},cu} = 4 \, (14 + \rho) \, \Oq_2 - 2 \, (1 + 2 \rho) \, \Oq_1 \,.
\end{eqnarray}
Again, the remaining functions are obtained from the following replacements
\begin{equation}
\begin{tabular}{lr}
$ \ast \, A^{{\rm WE}, cu}_{m, \, n} = 0$\,,
& 
$
\left\{
\begin{array}{l}
     m = 1, \ldots, \phantom{1}4, \,\, n = \phantom{1}5, \ldots 14\,, \\
     m = 5, \ldots, 10, \,\, n = 15, \ldots 20\,, \\
\end{array}
\right.
$
\\[5mm]
$\ast \, A^{{\rm WE},cu}_{2m-1, \, 2n-1} = A^{{\rm WE},cu}_{2m - 1, \,2n} 
\Big|_{\Oq^{(\prime)}_i \to \left(\frac{\Oq^{(\prime)}_i}{N_c} + 2 \, \Oq^{(\prime)}_{i+2}\right)}$
&
$m = 1, \ldots, 5\,, \,\,  n = 1, \ldots, 10\,,$
\\[7mm]
$\ast \, A^{{\rm WE},cu}_{2m, \, 2n} = N_c \, A^{{\rm WE},cu}_{2m - 1, \, 2n}\,, \quad A^{{\rm WE},cu}_{2m, \, 2n-1} = A^{{\rm WE},cu}_{2m-1, \, 2n}\,,$ \qquad
& 
$m = 1, \ldots, 5\,, \,\,  n = 1, \ldots, 10\,,$
\\[5mm]
$ \ast \,  A^{{\rm WE},cu}_{m, \, n} =  A^{{\rm WE},cu}_{n, m}$\,, \qquad
& 
$
\left\{
\begin{array}{l}
     m = 1, \ldots, 10, \,\, n = \phantom{1}1,  \ldots, 10 \,, \\
     m = 1, \ldots, 10, \,\, n = 11, \ldots, 20 \,, 
\end{array}
\right.
$
\\[6mm]
$\ast \, A^{{\rm WE},cu}_{m, \,n} = A^{{\rm WE},cu}_{m-10, \, n+10} \Big|_{\Oq_i \leftrightarrow \Oqp_i}$\,,
& 
$m = 11, \ldots, 20\,,  n = 1, \, \ldots, 10\,, $
\\[6mm]
$\ast \, A^{{\rm WE},cu}_{m, \,n} = A^{{\rm WE},cu}_{m-10, \, n-10} \Big|_{\Oq_i \leftrightarrow \Oqp_i}$\,,
& 
$m, n = 11, \, \ldots, 20 $\,.
\end{tabular}
\label{eq:repl-WE-cu}
\end{equation}
Note that in the case of one massive quark running in the loop, only half of the entries of the $20 \times 20$ matrix $\hat A^{{\rm X}, cq}$ are non zero, as the remaining contributions vanish due to the chiral structure of the corresponding massless line. For completeness and also to cross-check our calculation, we have recomputed the contribution to the WE diagram with two internal charm quarks, and found full agreement with the results of Ref.~\cite{Jager:2019bgk}, where these have been originally obtained in the case of NP in $b \to c \bar c s$ transitions. 
{ Since in Ref.~\cite{Jager:2019bgk} the results are expressed already in terms of the Bag parameters, i.e.\ by taking the matrix elements of the dimension-six four-quark operators between $B$-meson states, 
in Appendix~\ref{app:A} we present more general expressions for the contribution of $b \to c \bar c  d (s)$ transitions to WE, namely in an operator form, which has the advantage of being easily applicable also 
to other BSM studies, like
in the case of NP effects to baryon lifetimes.}
\\
Finally, we discuss the parametrisation of the matrix element of the $\Delta B = 0$ four-quark operators. We stress that in order to be consistent with the SM prediction of $\tau(B^+)/\tau(B_d)$, obtained in Ref.~\cite{Lenz:2022rbq}, and which we use in our analysis, we also parametrise the operators in HQET. In fact, any difference between operators defined in QCD or HQET arises only at dimension-seven, which we do not include in the present work. We thus have
\begin{equation}
\langle B | \Oq_i^{(\prime)} | B \rangle 
= f_B^2 \, m_B^2 \, \B_i^{(\prime)} \, ,
\label{eq:Par-ME}
\end{equation} 
where $f_B$ is the QCD decay constant, 
and $\B_i^{(\prime)}$ denote the corresponding Bag parameters. Note also that $\Bp_i = \B_i$, as it follows from parity conservation in QCD. 
Within vacuum insertion approximation (VIA), it is easy to show that 
\begin{equation}
\B_1 = \B_2 = 1\,, \qquad \B_5 = \B_6 = -1\,, \qquad 
\B_3 = \B_4 = \B_7 = \B_8 = 0\,.
\label{eq:Bag-VIA}
\end{equation}
We emphasise that for the Bag papameters $\B_{i}\,,$ with $i= 1, \ldots, 4$, also computations based on HQET sum rule are available~\cite{Kirk:2017juj, King:2021jsq}, however, the deviation from their VIA values is found to be small, at most of the order of few percents. Thus, in our numerical analysis, we use for~$\B_{i}\,,$ with $i= 1, \ldots, 4$, the most recent determination from Ref.~\cite{King:2021jsq}, but we fix the remaining Bag parameters to their VIA values. Deviations from the VIA assumption are accounted for in our error budget. 

\subsection{\boldmath BSM effects to mixing}
\label{asl}
Neutral $B$ mesons mix with their antiparticles, and the corresponding mass eigenstates are obtained by diagonalising the two-dimensional Hamiltonian matrix describing the evolution of the two-particle system $B_q - \bar B_q$. The mass and the decay width difference between the mass eigenstates, $\Delta M_q$ and $\Delta \Gamma_q$,
define the mixing observables, which are given by, see e.g.
Refs.~\cite{Buras:1984pq,Proceedings:2001rdi}
\begin{equation}
    \Delta M_q = 2 \, |M^q_{12}|\,, \qquad
    \Delta \Gamma_q = 2  \, |\Gamma^q_{12}| \cos{\phi_q}\,,
    \qquad 
    \phi_q = \arg\left(- M_{12}^q/\Gamma_{12}^q \right)\,,
\end{equation}
where $M^q_{12}$ and $\Gamma^q_{12}$ correspond respectively to the dispersive and absorptive part of the $\bar B_q \to B_q$ amplitude. Moreover, the semileptonic asymmetries read $a_{sl}^q = {\rm Im}(\Gamma^q_{12}/M^q_{12})$, cf. Eq.~\eqref{eq:asl-q}.
As BSM contributions in $b \to c \bar u d(s)$ have a negligible effect to $M^q_{12}$, see e.g. Ref.~\cite{Bobeth:2014rda}, in the following we only need to consider $\Gamma_{12}^q$.
The latter can be computed, similarly to Eq.~\eqref{eq:GammaB}, from
\begin{equation}
    \Gamma_{12}^q = \frac{1}{2 m_{B_q}} {\rm Im} \langle  B_q | {\cal T}_{12}| \bar B_q\rangle\,,
\end{equation}
where the transition operator ${\cal T}_{12}$ is formally defined as in Eq.~\eqref{eq:T-operator}, however we have added the suffix $12$ in order to stress that, contrary to the case of the total decay width, now the dimension-six matching coefficients  must be extracted by taking the matrix element of ${\cal T}_{12}$ between external $b$-quark states, describing respectively an incoming particle and an outgoing antiparticle. By factoring out the corresponding CKM elements, ${\cal T}_{12}$ can be decomposed as
\begin{equation}
    {\cal T}_{12} = \lambda_c^2 \, {\cal T}_{12}^{cc} + 2 \lambda_c \lambda_u \, {\cal T}_{12}^{cu} + \lambda_u^2 \,
    {\cal T}_{12}^{uu}\,,
\end{equation}
and here we denote $\lambda_x = V_{xb} V_{xq}^*$, while the upper labels indicate the internal quarks in the respective mixing diagrams, cf.\ Figure~\ref{fig:mixing-NP}. In the case of BSM effects to the $b \to c \bar u d (s)$ channels, the contribution of the Hamiltonian in Eq.~\eqref{eq:Heff-NP} to the imaginary part of the transition operator ${\cal T}_{12}^{cu}$, at LO-QCD and at dimension-six, can be thus presented as
\begin{equation}
{\rm Im} {\cal T}^{cu}_{12 , {\rm NP}} = \frac{G_F^2 m_b^2}{24 \pi} (1-\rho)^2
\left[
\sum_{m=1}^{2}\sum_{n = 1}^{20} C^{\rm SM}_m \, C_n \, M^{cu}_{mn} 
- \sum_{m,n = 1}^{2}
C_m^{\rm SM} C_n^{\rm SM} \, M^{cu}_{mn} \,\,
\right],
\label{eq:T12-mixing}
\end{equation}
where the functions $M^{cu}_{mn}$ are linear combinations of the $\Delta B = 2$ operators ${\cal Q}_i$, ${\cal Q}_i^\prime$, with the prime again denoting the exchange $P_L \leftrightarrow P_R$, i.e. 
\begin{align}
\Q_1 & 
= 4 \, (\bar q \, \gamma_\mu P_L \, b) \, (\bar q \gamma^\mu P_L \, b)\,,
&
\Q_2 & 
= 4 \, (\bar q P_R \, b) \, (\bar q P_R \, b)\,,
\label{eq:Q1-Q2-mix}
\\[1mm] 
\Q_3 & 
= 4 \, (\bar q \, \gamma_\mu P_L \, b) \, (\bar q \gamma^\mu P_R \, b)\,,
&
\Q_4 & 
= 4 \, (\bar q P_R \, b) \, (\bar q P_L \, b)\,,
\label{eq:Q3-Q4-mix}
\\[1mm]
\Q_1^\prime & 
= 4 \, (\bar q \, \gamma_\mu P_R \, b) \, (\bar q \gamma^\mu P_R \, b)\,, 
&
\Q_2^\prime & 
= 4 \, (\bar q P_L \, b) \, (\bar q P_L \, b)\,.
\label{eq:Q1p-Q2p-mix}
\end{align}
Note that ${\cal Q}_{3,4}$ coincide with the corresponding primed operators, and that again, for the sake of a cleaner notation, we omit to include a label $q$ for the respective operators. However, this should not cause any confusion, since the external light quark in Eqs.~\eqref{eq:Q1-Q2-mix} - \eqref{eq:Q1p-Q2p-mix}, always corresponds to the spectator quark in the $B_q$ meson.  
We also emphasise that in deriving the above basis we have made use of Fierz transformations and applied the equations of motion for the $b$-quark $u$- and $v$-spinors, in order to reduce the number of independent operators required at dimension-six 
\footnote{In fact, the most general $\Delta B = 2$ basis contains 8 independent operators, see e.g. Ref.~\cite{Gabbiani:1996hi}.}. Specifically, this allows us to rewrite the corresponding colour-rearranged operators $\tilde {\cal Q}_i, \tilde {\cal Q}^\prime_i$, in terms of a linear combinations of the ones in  Eqs.~\eqref{eq:Q1-Q2-mix} - \eqref{eq:Q1p-Q2p-mix}, always up to $1/m_b$ corrections~\footnote{Note, that there is some arbitrariness in the choice of the operator basis, since Fierz transformations can always be used to replace an operator with a different one. We have chosen to express the $\Delta B = 2$ operators in terms of the colour-singlet ones only. However, alternative choices can be found in the literature.}. Namely
\begin{align}
    & \tilde {\cal Q}_1  = {\cal Q}_1\,, \qquad \tilde {\cal Q}_1^\prime  = {\cal Q}_1^\prime \,, 
    \\[1.5mm]
    & \tilde {\cal Q}_2 = - \frac12 {\cal Q}_1 - {\cal Q}_2 + {\cal O}(1/m_b)\,, 
    \\[1.5mm]
    &  \tilde {\cal Q}_2^\prime  = - \frac12 {\cal Q}_1^\prime - {\cal Q}_2^\prime + {\cal O}(1/m_b)\,,
    \\[1.5mm]
    &  \tilde {\cal Q}_3  = -2 {\cal Q}_4\,, \qquad \tilde {\cal Q}_4  = - \frac{1}{2} {\cal Q}_3\,.
\end{align}
It is worth stressing that $\Gamma_{12}^{q, cu}$ originates from the interference of $b \to c \bar u q$ and $b \to u \bar c q$ transitions and is computed from the imaginary part of the mixing diagram in Figure~\ref{fig:mixing-NP}. As we only consider NP in the $b \to c \bar u d (s)$ channels, 
BSM contributions to $\Gamma_{12}^{q, cu}$ correspond to the time-ordered product of the BSM Hamiltonian describing the $b \to c \bar u d (s)$ decays with the SM one describing $b \to u \bar c d(s)$.
Thus, in Eq.~\eqref{eq:T12-mixing}, only 
the first two rows of the general 20 $\times$ 20 matrix $M_{mn}^{cu}$, that we would obtain by extending the BSM Hamiltonian in Eq.~\eqref{eq:Heff-NP} to include also the set of 20 operators for the $b \to u \bar c d (s)$ transitions, actually contribute. However, for completeness, here we present the full coefficient matrix $\hat M^{cu}$,
with the remark that only the functions $M_{mn}^{cu}$ with $m = 1,2,$ and $n = 1,\, \ldots , 20$, actually arise in the specific NP scenario we are considering. This matrix reads
\begin{eqnarray}
&\displaystyle
M^{cu}_{1,1} 
= 2 \, (1 + 2 \rho) \, {\cal Q}_2 - (1 - \rho) \, {\cal Q}_1\,, 
\qquad
M^{cu}_{1,3} 
= - 3 \sqrt{\rho} \, ({\cal Q}_1 + 2 {\cal Q}_2)\,,
&
\nonumber \\[2mm]
& \displaystyle 
M^{cu}_{1,15} 
= - \frac 1 2 ((1 + 2 \rho) {\cal Q}_3 + 2 (2 + \rho) {\cal Q}_4)\,, 
\qquad
M^{cu}_{1,17} 
= - \frac 3 2 \sqrt{\rho} \, ( {\cal Q}_3 + 2 {\cal Q}_4)\,, 
&
\\[2mm]
& \displaystyle
M^{cu}_{1,19} 
= 6 \sqrt{\rho} \, (2 {\cal Q}_4 -  {\cal Q}_3)\,,
&
\nonumber
\end{eqnarray}
\vspace{1mm}
\begin{eqnarray}
& \displaystyle
M^{cu}_{3,5} = - \frac32 \sqrt{\rho} \left( {\cal Q}_1 + 2 {\cal Q}_2\right)\,, \qquad 
M^{cu}_{3,7} = - \frac32 ({\cal Q}_1 + 2 {\cal Q}_2)\,,
\quad
M^{cu}_{3,9} = - 18 ({\cal Q}_1 + 2 {\cal Q}_2)\,, 
&
\nonumber \\[2mm]
& \displaystyle
M^{cu}_{3,11} = 3 \sqrt{\rho} \, {\cal Q}_3 \,, \qquad
M^{cu}_{3,13} = - 6 {\cal Q}_3 \,,
&
\end{eqnarray}
\vspace{1mm}
\begin{eqnarray}
& \displaystyle
    M^{cu}_{5,5} = - \frac 1 4 \left((1-\rho) {\cal Q}_1 - 2 (1+ 2 \rho) {\cal Q}_2\right)\,, \qquad 
    M^{cu}_{5,7} =  \frac 3 2 \sqrt{\rho} \, {\cal Q}_2\,,
    \qquad
    M^{cu}_{5,9} =  6 \sqrt{\rho}  \, ({\cal Q}_1 +  {\cal Q}_2)\,, 
    &
    \nonumber \\[2mm]
& \displaystyle
    M^{cu}_{5,11} = - \frac 1 2 ((1 + 2 \rho) {\cal Q}_3 + 2 (2 + \rho) {\cal Q}_4) \,,
    \quad
    M^{cu}_{5,13} = \frac 3 2 \sqrt{\rho} \, {\cal Q}_3 \,,
\end{eqnarray}
\vspace{1mm}
\begin{eqnarray}
& \displaystyle
    M^{cu}_{7,1} =\frac 3 2 \sqrt{\rho} \, {\cal Q}_1 \,, \qquad 
    M^{cu}_{7,3} = - \frac 3 2 ({\cal Q}_1 + 2 {\cal Q}_2)\,,
    \qquad
    M^{cu}_{7,15} =\frac 3 2 \sqrt{\rho} \, {\cal Q}_4 \,, 
&
\nonumber \\[2mm]
& \displaystyle    
M^{cu}_{7,17} = - \frac14 ((1 - \rho)  {\cal Q}_3 - 2 (1+ 2\rho) {\cal Q}_4) \,, \quad
M^{cu}_{7,19} = - ( (5 + \rho)  {\cal Q}_3 + 2 (1+ 2 \rho) {\cal Q}_4)\,,
&
\end{eqnarray}
\vspace{1mm}
\begin{eqnarray}
& \displaystyle
    M^{cu}_{9,1} = 6 \sqrt{\rho} \, ({\cal Q}_1 + 4 {\cal Q}_2)\,, \qquad 
    M^{cu}_{9,3} = - 18 ({\cal Q}_1 + 2 {\cal Q}_2)\,,
    \qquad
    M^{cu}_{9,15} = - 6 \sqrt{\rho} \, ({\cal Q}_3 + {\cal Q}_4) \,,
    \quad 
&
\nonumber \\[3mm]
& \displaystyle
M^{cu}_{9,17} = - ( (5 + \rho) {\cal Q}_3 + 2 (1+ 2\rho) {\cal Q}_4) \,,
\quad
M^{cu}_{9,19} = - 4 ( (13 - \rho)  {\cal Q}_3 - 2 (1+ 2 \rho) {\cal Q}_4) \,. \quad
&
\end{eqnarray}
\vspace{1mm}
The remaining entries can be obtained as follows:
\begin{equation}
\begin{tabular}{lr}
$\ast \, M^{cu}_{2m - 1, \, 2n} =M^{cu}_{2m , \, 2n - 1} =   M^{cu}_{2m-1 , \, 2n-1}\Big|_{{\cal Q}_i \to \tilde {\cal Q}_i} \,, $
& $m = 1, \ldots 5\,, n = 1, \ldots, 10\,,$  
\\[4mm]
$\ast \, M^{cu}_{2m, \, 2n} = N_c \, M^{cu}_{2m-1 , \, 2n-1}\Big|_{{\cal Q}_i \to \tilde {\cal Q}_i} \,, $
& $m = 1, \ldots 5\,, n = 1, \ldots, 10\,,$  
\\[4mm]
$\ast \, M^{cu}_{m, \, n} =  M^{cu}_{m - 10 , \, n+ 10}\Big|_{{\cal Q}_i \to {\cal Q}_i^\prime} \,, $
& $m = 11, \ldots 20\,, n = 1, \ldots, 10\,,$  
\\[4mm]
$\ast \,  M^{cu}_{m, \, n} =  M^{cu}_{m - 10 , \, n- 10}\Big|_{{\cal Q}_i \to {\cal Q}_i^\prime} \,, $
& $m, n = 11, \ldots 20\,.$
\end{tabular}
\end{equation}
For completeness and also as a cross-check of our calculation, we have recomputed the contribution to the mixing diagram in Figure~\ref{fig:mixing-NP} with two internal charm quarks. Again, these coefficients were firstly derived in Ref.~\cite{Jager:2019bgk} for the case of NP in $b \to c \bar c s$ transitions. However, we have found that in Ref.~\cite{Jager:2019bgk} the mixed contributions from $Q_i \otimes Q_i^\prime$ operators have been overlooked, leading to incomplete results with half of the coefficients incorrectly neglected. Therefore, even if not relevant under our assumption of NP in $b\to c \bar u d (s)$ decays, we present here the complete expressions for the NP contribution to the imaginary part of ${\cal T}_{12}^{cc}$ \footnote{Note, that Eq.~\eqref{eq:T12-mixing-cc} follows from the BSM Hamiltonian in Eq.~\eqref{eq:Heff-NP} with the replacement of $ u \to c$.}:
\begin{equation}
{\rm Im} {\cal T}^{cc}_{12 , {\rm NP}} =
\frac{G_F^2 m_b^2}{24 \pi} \sqrt{1- 4 \rho}
\left[
\sum_{m, n = 1}^{20} C_m \, C_n \, M^{cc}_{mn} 
- \sum_{m,n = 1}^{2}
C_m^{\rm SM} C_n^{\rm SM} \, M^{cc}_{mn} \,\,
\right]\,,
\label{eq:T12-mixing-cc}
\end{equation}
with
\begin{eqnarray}
    & \displaystyle
    M_{1,1}^{cc} = - ((1- 4 \rho) {\cal Q}_1 - 2 (1+ 2 \rho) {\cal Q}_2)\,, \quad
    M_{1,3}^{cc} = - 3 \sqrt{\rho} \, ({\cal Q}_1 + 2 {\cal Q}_2)\,, \quad 
    M_{1,5}^{cc} = 3 \rho \, {\cal Q}_1\,, 
    &
    \nonumber \\[2mm]
    & \displaystyle
    M_{1,7}^{cc} = \frac32 \sqrt{\rho} \, {\cal Q}_1\,, \quad
    M_{1,9}^{cc} = 6 \sqrt{\rho} \, ({\cal Q}_1 + 4 {\cal Q}_2)\,, \quad 
    M_{1,11}^{cc} = 12 \rho \, {\cal Q}_4 \,, 
    &
    \nonumber \\[2mm]
    & \displaystyle
    M_{1,13}^{cc} = 3 \sqrt{\rho} \, {\cal Q}_3\,, \quad
    M_{1,15}^{cc} = - \frac12 (4 (1- \rho) {\cal Q}_4 + (1+ 2 \rho) {\cal Q}_3)\,, 
    &
    \nonumber \\[2mm]
    & \displaystyle
    M_{1,17}^{cc} = - \frac32 \sqrt{\rho} \, ({\cal Q}_3+ 2 {\cal Q}_4)\,,
    \quad
    M_{1,19}^{cc} = 6 \sqrt{\rho} \, (2 {\cal Q}_4 - {\cal Q}_3)\,,
    &
\end{eqnarray}
\vspace{1mm}
\begin{eqnarray}
& \displaystyle
    M_{3,3}^{cc} = 12 \rho \,  ({\cal Q}_1 + 2  {\cal Q}_2)\,, \quad
    M_{3,5}^{cc} = - \frac 3 2  \sqrt \rho \, ({\cal Q}_1 + 2 {\cal Q}_2)\,, \quad 
    M_{3,7}^{cc} = -\frac 3 2 (1- 2 \rho) \, ({\cal Q}_1+ 2 {\cal Q}_2)\,,
    &
    \nonumber \\[2mm]
    & \displaystyle
    M_{3,9}^{cc} = -18 (1- 2 \rho) ({\cal Q}_1 + 2 {\cal Q}_2)\,, \quad
    M_{3,13}^{cc} =- 6 (1- 2 \rho) {\cal Q}_3 \,, \quad 
    M_{3,15}^{cc} = \frac 3 2 \sqrt{\rho} \, {\cal Q}_3 \,,
    &
    \nonumber \\[2mm]
    & \displaystyle
    M_{3,17}^{cc} = 3 \rho \,  {\cal Q}_3\,,\quad 
    M_{3,19}^{cc} = 36 \rho \,  {\cal Q}_3\,, 
    &
\end{eqnarray}
\vspace{1mm}
\begin{eqnarray}
& \displaystyle
    M_{5,5}^{cc} = - \frac14  ((1- 4 \rho) {\cal Q}_1 - 2 (1+ 2 \rho) {\cal Q}_2)\,, \quad
    M_{5,7}^{cc} = \frac32 \sqrt{\rho} \, {\cal Q}_2\,, \quad 
    M_{5,9}^{cc} = 6 \sqrt{\rho} \,  ({\cal Q}_1 + {\cal Q}_2)\,,
    &
    \nonumber \\[2mm]
    & \displaystyle
    M_{5,15}^{cc} = 3 \rho \, {\cal Q}_4 \,, \quad
    M_{5,17}^{cc} =  \frac 3 2 \sqrt{\rho} \,  {\cal Q}_4 \,, \quad 
    M_{5,19}^{cc} = - 6 \sqrt{\rho} \,  ( {\cal Q}_3 + {\cal Q}_4 ) \,,
    &
\end{eqnarray}
\vspace{1mm}
\begin{eqnarray}
& \displaystyle
    M_{7,7}^{cc} = 3 \rho \,  {\cal Q}_2 \,, \quad
    M_{7,9}^{cc} = 12 \rho \, ({\cal Q}_1 + {\cal Q}_2)\,, \quad
    M_{7,17}^{cc} =- \frac 1 4  ((1 - 4 \rho) {\cal Q}_3 -2 (1+ 2 \rho) {\cal Q}_4)\,,
    &
    \nonumber \\[2mm]
    & \displaystyle
    M_{7,19}^{cc} = - ((5 - 8 \rho) {\cal Q}_3 + 2 (1 + 2 \rho) {\cal Q}_4))\,,
    &
\end{eqnarray}
\vspace{1mm}
\begin{eqnarray}
& \displaystyle
    M_{9,9}^{cc} = 48 \rho \,  (2 {\cal Q}_1 + 5 {\cal Q}_2)\,, \quad
    M_{9,19}^{cc} = - 4 ((13 - 28 \rho) {\cal Q}_3 - 2 (1+ 2 \rho) {\cal Q}_4)\,. 
    &
\end{eqnarray}
\vspace{1mm}
The remaining entries can be obtained as follows
\begin{equation}
\begin{tabular}{lr}
$ \ast \, M^{cc}_{2m - 1, \, 2n} = M^{cc}_{2m , \,2n - 1} = M^{cc}_{2m-1 , \,2n - 1}\Big|_{{\cal Q}_i \to \tilde {\cal Q}_i}$
&
$m = 1, \ldots, 5\,, \,\,  n = 1, \ldots, 10\,,$
\\[4mm]
$\ast \, M^{cc}_{2m, \, 2n}  = N_c \, M^{cc}_{2m - 1, \, 2n- 1}\Big|_{{\cal Q}_i \to \tilde {\cal Q}_i}$
\qquad
& 
$m = 1, \ldots, 5\,, \,\,  n = 1, \ldots, 10\,,$
\\[3mm]
$\ast \, M^{cc}_{m, \, n} =  M^{cc}_{n, m}$\,, \qquad
& 
$
\left\{
\begin{array}{l}
     m = 1, \ldots, 10\,, \, n = \phantom{1}1, \ldots, 10\,,\\
     m = 1, \ldots, 10, \,\, n = 11, \ldots, 20 \,, 
\end{array}
\right.
$
\\[6mm]
$\ast \, M^{cc}_{m, \,n} = M^{cc}_{m-10, \, n+10}\Big|_{{\cal Q}_i \to {\cal Q}^\prime_i}$\,,
& 
$m = 11, \ldots, 20\,,  n = 1, \, \ldots, 10\,, $
\\[3mm]
$\ast \, M^{cc}_{m, \,n} = M^{cc}_{m-10, \, n-10}\Big|_{{\cal Q}_i \to {\cal Q}^\prime_i}$\,,
& 
$m, n = 11, \, \ldots, 20 $\,.
\end{tabular}
\label{eq:repl-WE}
\end{equation}
Finally, we discuss the parametrisation of the matrix element of the $\Delta B = 2$ operators defined in Eqs.~\eqref{eq:Q1-Q2-mix} - \eqref{eq:Q1p-Q2p-mix}. Again, in order to be consistent with the SM prediction of $a_{sl}^q$ and $\Gamma^q_{12}$ obtained in Ref.~\cite{Lenz:2019lvd}, and which we use in our analysis, we also parametrise the operators in full QCD. It is easy to show that in this case
\begin{align}
&\langle B_q| \Q_1 | \bar B_q \rangle =
\frac{8}{3} f_{B_q}^2 m_{B_q}^2 \, \BM_1^q \,,
\label{eq:Q1-par}
\\[1mm]
&\langle B_q| \Q_2 | \bar B_q \rangle  =
- \frac{5}{3} f_{B_q}^2 m_{B_q}^2 \frac{m_{B_q}^2}{(m_b + m_q)^2} \, \BM_2^q \,,
\\[1mm]
&\langle B_q| \Q_3 | \bar B_q \rangle = 
- f_{B_q}^2 m_{B_q}^2 \left(2 + \frac 4 3 \frac{m_{B_q}^2}{(m_b+ m_q)^2} \right) \BM_3^q \, ,
\\[1mm]
& \langle B_q| \Q_4 | \bar B_q \rangle = 
f_{B_q}^2 m_{B_q}^2 \left(\frac 1 3 + 2 \frac{m_{B_q}^2}{(m_b+ m_q)^2} \right) \BM_4^q\,,
\label{eq:Q4-par}
\end{align}
with $m_q$ being the mass of the light quark $q = d, s$, while, due to parity conservation in QCD, the matrix element of ${\cal Q}_{1,2}^\prime$ have exactly the same parametrisation of the corresponding unprimed operators.
The Bag parameters $\BM^q_i$ in Eqs.~\eqref{eq:Q1-par} - \eqref{eq:Q4-par} have been determined using both Lattice QCD~\cite{FermilabLattice:2016ipl,Dowdall:2019bea} 
and HQET sum rules \cite{Kirk:2017juj, King:2019lal} methods. 
In our numerical analysis we use the average values presented explicitly in Ref.~\cite{DiLuzio:2019jyq}.

\section{Numerical analysis}
\label{Numerics} 
In this section we discuss our analysis and the corresponding bounds on the NP Wilson coefficients of the effective Hamiltonian 
given in Eq.~(\ref{eq:Heff-NP}), derived from the study of the lifetime ratio $\tud$ and of the mixing observables $a_{sl}^q$ and $\Delta \Gamma_q$, under the assumption of BSM effects only in the decay channels $b \to c \bar{u} d(s)$. In addition, we also compare 
our allowed regions
with the ones obtained in 
Ref.~\cite{Cai:2021mlt}, originating from the analysis of the hadronic decays 
$\bar{B}_{(s)}^0 \to D_{(s)}^{(*)+} L^-$, with $L \in \{ \pi, \rho, K^{(*)}\}$.

\subsection{Strategy}
The contours for the NP Wilson coefficients $C_i^{(\prime)}$ are determined by introducing the $\chi^2$-distribution function
\begin{equation}
\chi^2(f; \, c_1, \ldots, c_n) = \left(\frac{f^{\rm exp} - f^{\rm SM} - f^{\rm NP}(c_1, \ldots, c_n)}{\sqrt{\sigma_{\rm exp}^2 + \sigma_{\rm th}^2}}\right)^2, 
\label{eq:chi-sq}
\end{equation}
where 
$f \in \{\tud, \, a_{\rm sl}^d, \, a_{\rm sl}^s, \Delta \Gamma_d, \Delta \Gamma_s\}$ denotes the  observable considered, with the NP contribution depending on $n$-degrees of freedom $c_i$
(in our case the specific NP Wilson coefficients we want to constrain), while
$\sigma_{\rm th}$ and $\sigma_{\rm exp}$ indicate respectively
the theoretical and the experimental uncertainties.
Note, that the correlation between the theoretical and experimental values is neglected.
The allowed regions at a given confidence level are then obtained by demanding $\Delta \chi^2 \leq \chi^2_{n}$, where
$\Delta \chi^2 = \chi^2 - \chi^2_{\rm min}$ 
and $ \chi^2_2 = 2.30, \, 6.18, \ldots, $ in correspondence respectively to the $1\sigma$, $2\sigma, \ldots,$ intervals. 
\\
Our results are presented as 
$2\sigma$-favoured regions for the NP Wilson coefficients at the scale $\mu = m_b$, and always in the case of two non-vanishing NP operators at the time. 
In fact, at this scale, the single NP Wilson coefficient scenario looks quite unrealistic as the operators mix under renormalisation when running from the higher NP scale down to lower scale $\mu = m_b$. 
Moreover, as the evolution matrix is block-diagonal i.e.\ $\hat U = {\rm diag} (\hat a_2, \hat b_2, \hat c_2, \hat d_4)$ with $\hat a_n, \ldots, \hat d_n,$ being $n \times n$ mixing matrices, see e.g.\ Ref.~\cite{Ciuchini:1997bw}, the mixing occurs only between
$Q_1$ and $Q_2$, $Q_3$ and $Q_4$, etc..
\\
For consistency, all input used in our analysis, for a given observable, coincide with those adopted in the corresponding SM predictions. Specifically, we follow Ref.~\cite{Lenz:2022rbq} in the case of $\tau(B^+)/\tau(B_d)$ and Ref.~\cite{Lenz:2019lvd} for the mixing observables. 
For this reason, we do not show here any list of input, but we refer to the corresponding references for the explicit values.
Concerning the additional Bag parameters, which do not emerge in the SM, 
we adopt, for the case of the lifetime ratio,  their VIA values, i.e.\ ${\cal B}_{5,6} = -1$ and 
${\cal B}_{7,8} = 0$, while for the mixing observables we use the available averaged results
\cite{DiLuzio:2019jyq} based on both HQET sum rules and Lattice QCD determinations. 
\\
Finally, a comment on the treatment of the uncertainties from the NP contributions.
In the case of the lifetime ratio $\tau(B^+)/\tau(B_d)$, the total uncertainty is dominated by the theoretical one. As stated previously, in our analysis we include only the NP effects at LO-QCD and at order~$1/m_b^3$. However, as $\alpha_s$ corrections have a sizeable effect in the corresponding SM prediction, mainly due to $\Gamma(B^+)$, we add in quadrature to $\sigma_{\rm th}$ a conservative 30\%
uncertainty of the central value of the NP contribution, in order to account for missing higher-order QCD and power corrections, as well as for the deviation of the new Bag parameters ${\cal B}_i$, with $i = 5, \ldots, 8,$ from their VIA values. 
Conversely, the total uncertainty for $a_{sl}^d$ is currently strongly dominated by the experimental one, hence, in this case we do not add any additional theoretical uncertainty from the NP contribution. As for the remaining mixing observables, we find that, 
given the current SM and experimental uncertainties and in the specific BSM scenario we are considering, they do not provide any 
significant constraints, see the next subsection.
Hence, we do not  discuss further the treatment of the uncertainties in the case of $a_{sl}^s$ and $\Delta \Gamma_q$.
\subsection{Results}
Under the assumption of NP effects in $b \to c \bar{u} d(s)$ transitions, Figure~\ref{fig:combinations-mb-b-to-c-u-d-only} shows our bounds on the NP Wilson coefficients of the effective Hamiltonian Eq.~(\ref{eq:Heff-NP}), originating from $\tud$ and $a_{\rm fs}^d$. In fact, we find that $\Delta \Gamma_d$, $a_{\rm sl}^s$, and $\Delta \Gamma_s$, do not provide any additional constraints 
within reasonable regions of the NP parameter space.
This follows either from the weak experimental bounds or from the weak theoretical sensitivity, e.g.\ the contribution of the decay channel $b \to c \bar{u} s$
to $\Delta \Gamma_s$ is CKM suppressed.
The individual plots of Figure~\ref{fig:combinations-mb-b-to-c-u-d-only} correspond to the 2$\sigma$ regions for each of the NP Dirac structures, 
with the colour singlet and colour rearranged operators being respectively on the horizontal and on the vertical axis. 
For convenience, our results are presented 
at the scale $\mu = m_b$, in fact, they do not 
change dramatically when considering instead the scale $\mu = M_W$. For completeness, however, we show the corresponding results 
in Appendix~\ref{supplement}. Moreover, as explained in Section~\ref{asl}, in the case of NP in $b\to c \bar u d(s)$ transitions, only the operators $Q_{1}, \ldots, Q_4$, and $Q_5^\prime, \ldots, Q_{10}^{\prime}$ contribute to $a_{sl}^q$. In the corresponding plots, in order to better understand the impact that future experimental improvements would have on the resulting regions, 
in addition to the current experimental bound on $a_{sl}^d$, we show also the projection A given in Eq.~\eqref{eq:asl_HFLAV-proj}. Finally, to facilitate the comparison with the results obtained in Ref.~\cite{Cai:2021mlt}, 
in our plots we also include the favoured regions from the analysis of hadronic $B$-meson decays performed in the latter reference and here denoted as [CDLY'21].
\\
From Figure~\ref{fig:combinations-mb-b-to-c-u-d-only} we see that the lifetime ratio $\tud$ gives a cross-like region in the $C_1^{\rm NP}-C_2^{\rm NP}$ plane. Thus, considering this observable alone leaves the two Wilson coefficients $C_1^{\rm NP}$ and $C_2^{\rm NP}$ quite unconstrained. However, adding the bound from $a_{sl}^d$, we find, even with the current limited experimental knowledge, an interesting vertical region, so that the overlap of the constraints from 
$\tud$ and $a_{sl}^d$ yields a quite restricted area for the allowed values of $C_1^{\rm NP}$ and~$C_2^{\rm NP}$.
Including in addition the results of Ref.~\cite{Cai:2021mlt} from the analysis of the non-leptonic $B$-meson decays, we find a further constrained area, which is currently allowed. Future more precise experimental values for $a_{sl}^d$
will give us the possibility of excluding an overlap between all three observables -- in the case that $a_{sl}^d$ would stay at the current central experimental value (as indicated in the figure), or of confirming it, if the central value of $a_{sl}^d$
should move to the left in the first plot of Figure~\ref{fig:combinations-mb-b-to-c-u-d-only}. 
\\
In the $C_3^{\rm NP}-C_4^{\rm NP}$ plane the situation is a little worse, since here the mixing contribution is helicity suppressed and therefore we have less sensitivity to NP contributions.
However, taking into account the projection~B shown in Eq.~\eqref{eq:asl_HFLAV-proj}, instead of the projection~A, 
it would be possible to obtain
a stronger bound on the size of 
BSM effects in $C_3^{\rm NP}$ and $C_4^{\rm NP}$.
\\
Unfortunately, we do not get any mixing constraints in the 
$C_5^{\rm NP} - C_6^{\rm NP}$, $C_7^{\rm NP} - C_8^{\rm NP}$, 
$C_9^{\rm NP} - C_{10}^{\rm NP}$, $C_1^{\prime \rm NP} - C_2^{\prime \rm NP}$
and $C_3^{\prime \rm NP} - C_4^{\prime \rm NP}$ - planes, 
see Section \ref{asl}, therefore the corresponding Wilson coefficients 
are less constrained in these cases.
Finally, in the 
$C_5^{\prime \rm NP} - C_6^{\prime \rm NP}$,
$C_7^{\prime \rm NP} - C_8^{\prime \rm NP}$, and 
$C_9^{\prime \rm NP} - C_{10}^{\prime \rm NP}$ - planes, 
we again find complementary constraints from mixing and lifetimes.

\begin{longfigure}{c}
    \includegraphics[width=0.42\textwidth]{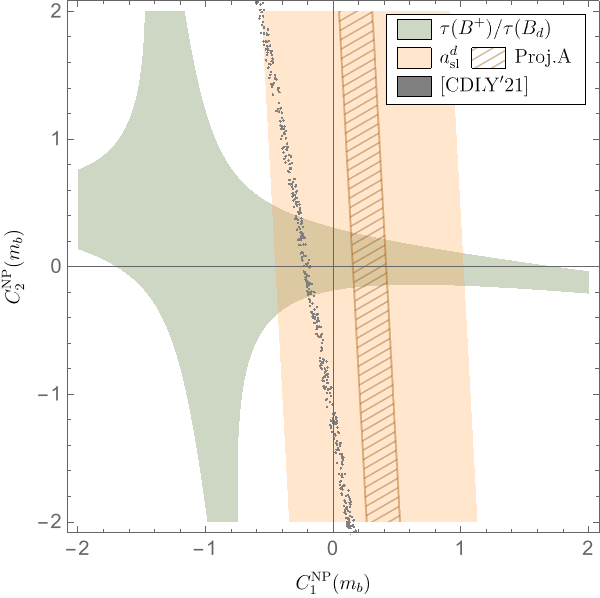}\qquad
    \includegraphics[width=0.42\textwidth]{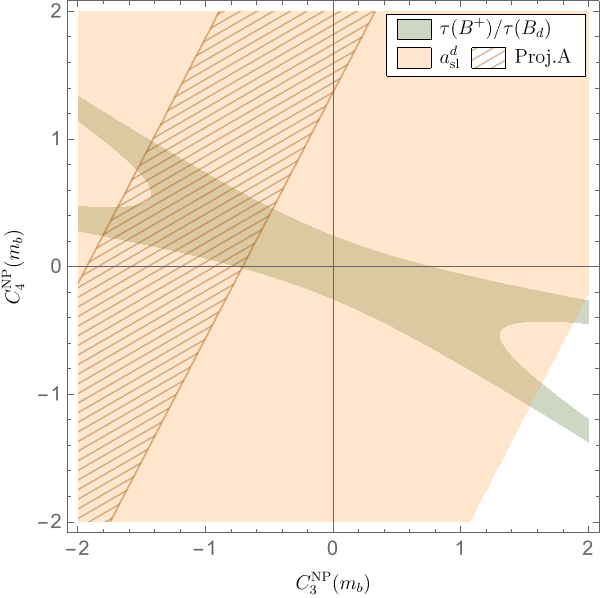} \\
    \includegraphics[width=0.42\textwidth]{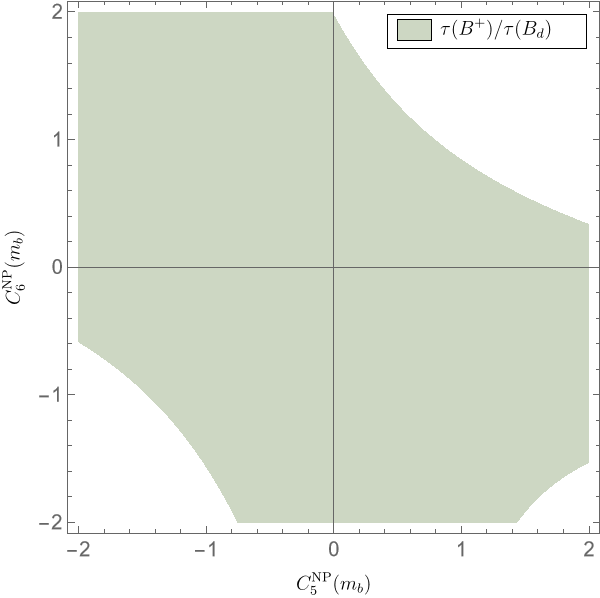}\qquad
    \includegraphics[width=0.42\textwidth]{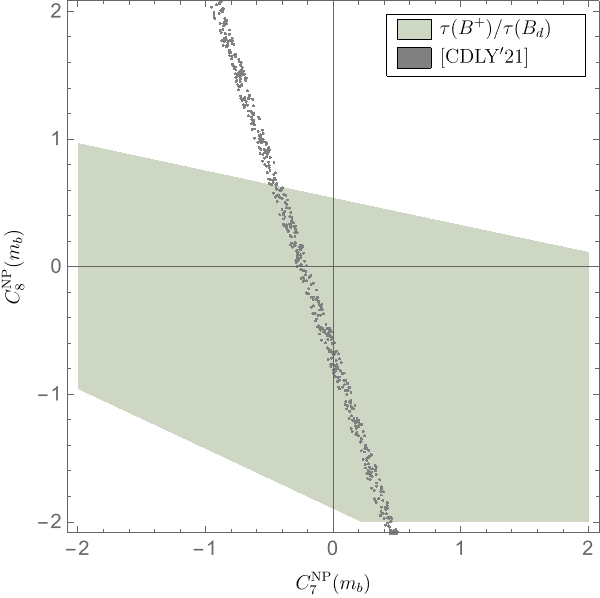} \\
    \includegraphics[width=0.42\textwidth]{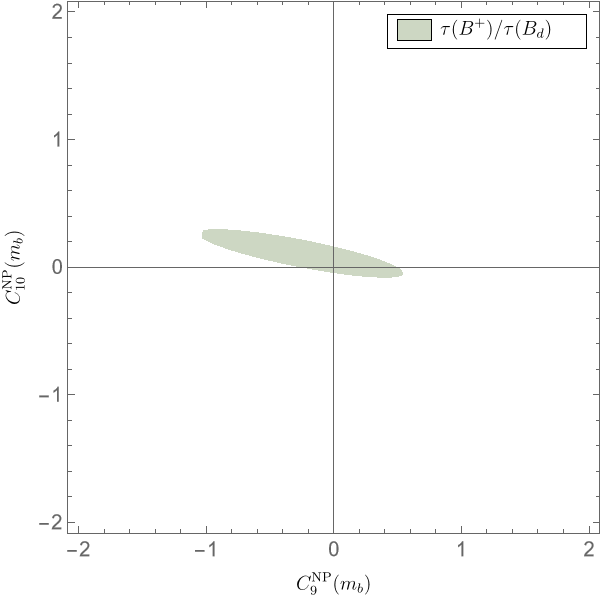}\qquad
    \includegraphics[width=0.42\textwidth]{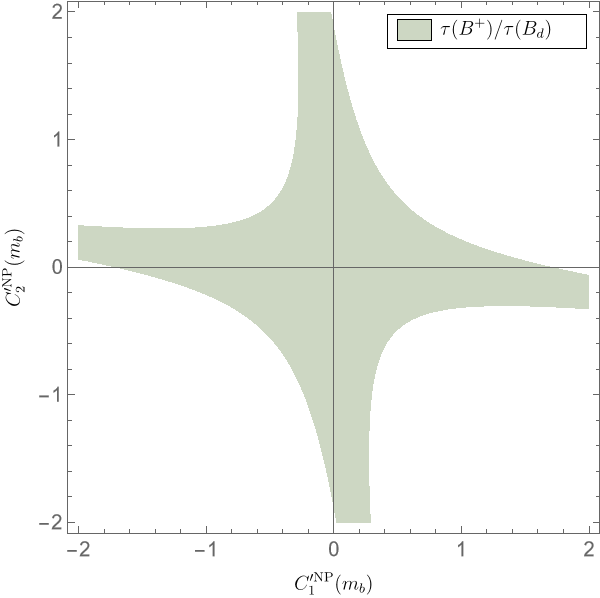} \\
    \includegraphics[width=0.42\textwidth]{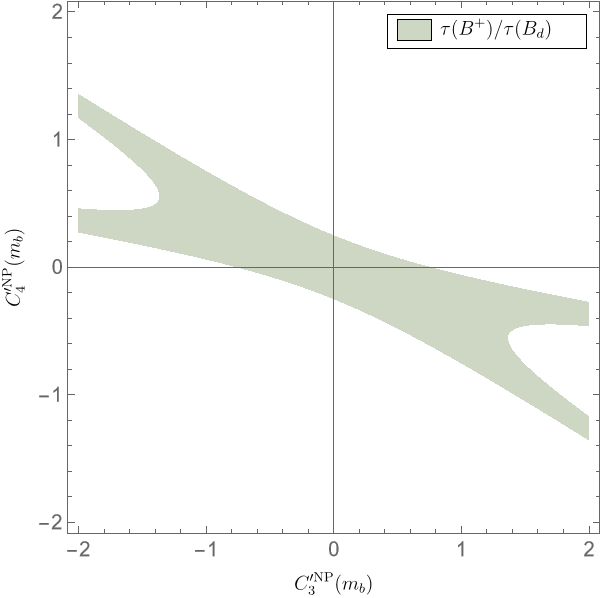}\qquad
    \includegraphics[width=0.42\textwidth]{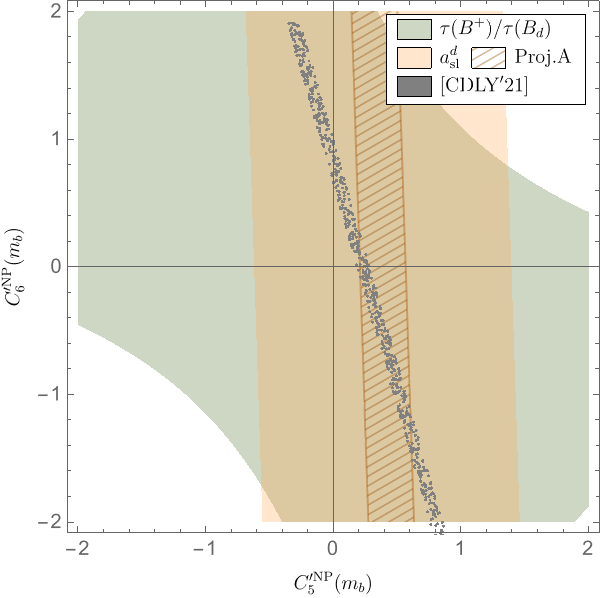} \\
    \includegraphics[width=0.42\textwidth]{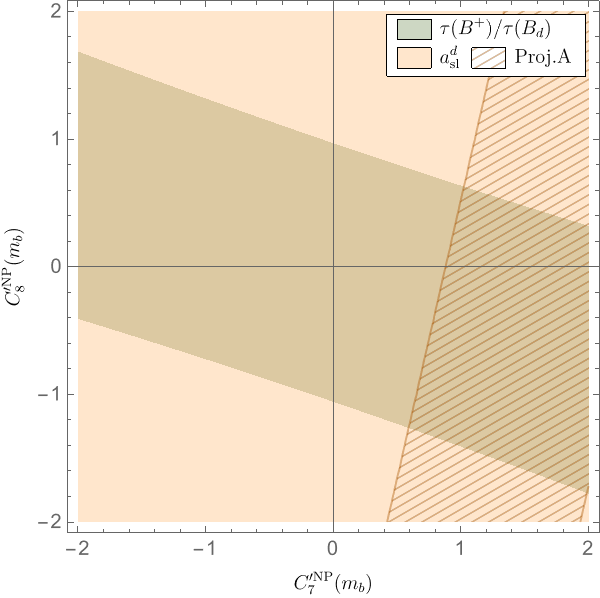}\qquad
    \includegraphics[width=0.42\textwidth]{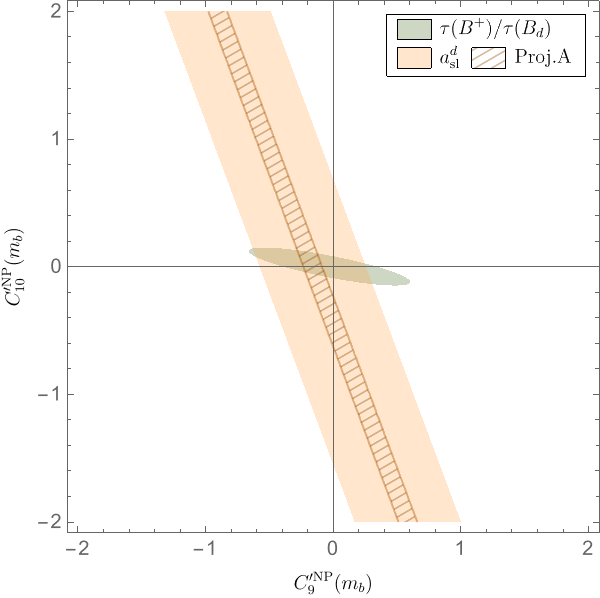} \\
    \caption{$2\sigma$ contours 
    for the NP Wilson coefficients $C_i^{(\prime) {\rm NP}}(m_b)$, 
    assuming BSM effects only in $b \to c \bar u d (s)$ transitions.}
    \label{fig:combinations-mb-b-to-c-u-d-only}
\end{longfigure}
\noindent

\section{Conclusion}
\label{Conclusion} 
In this work, we have computed the contribution of the 20
$\Delta B = 1$ four-quark operators, describing BSM effects in non-leptonic $b \to c \bar u d(s)$ transitions to lifetime and mixing observables. In this respect, 
our analytical results complement those obtained in Refs.~\cite{Jager:2019bgk, Jager:2017gal} for the case of the $b \to c \bar c s$ decay. 
However, we have found that in the case of mixing, half of the coefficients for the NP contributions from the $b \to c \bar c s$ channel had been wrongly neglected in Ref.~\cite{Jager:2019bgk}, leading to incomplete results. Therefore, in this paper, we have also presented the complete expressions for these contributions, as additional results.
We stress though, that we have not performed any comprehensive analysis of BSM effects in the $b \to c \bar c s$ transition, and we leave this for a future work.
\\
As a phenomenological application of our results, we have investigated the possible size
of BSM effects in $b \to c \bar u d (s)$ transitions and we have found that $\tud$ and $a_{sl}^d$ give complementary constraints on some of the NP Wilson coefficients.
On the other side, we have found no significant constraints coming from the other mixing observables, 
$\Delta \Gamma_{d,s}$ and $a_{sl}^d$, within reasonable ranges for the NP Wilson coefficients. Moreover, the BSM regions determined in Ref.~\cite{Cai:2021mlt}
to explain the discrepancies in the decays $\bar{B}_{(s)}^0
\to D_{(s)}^{(*)+} L^-$, with $L \in \{ \pi, \rho, K^{(*)}\}$, currently 
can not be completely excluded, although,
particularly in the case of $C_1^{\rm NP}$ and $C_2^{\rm NP}$, the allowed NP
parameter space can shrink dramatically, when all
three constraints from the lifetimes, mixing and
hadronic decays are taken into account. 
Future improvements in the experimental precision of the semileptonic CP asymmetries 
will considerably reduce the allowed regions for the NP Wilson coefficients, and our investigation of the experimental projections has shown that these improvements have the potential to exclude
or confirm a BSM scenario that explains the observed anomalies 
in the hadronic $B$-meson decays.
In this respect, we would like to emphasize that a measurement of the asymmetry $a_{sl}^s$
with the flavour-specific decay $\bar B_s \to D_s^+ \pi^-$, may provide an opportunity to unambiguously identify CP-violating BSM contributions to the decay channel $b \to c \bar{u} d$, see Refs.~\cite{Gershon:2021pnc, Fleischer:2016dqd}.
\\
In order to gain more insights on the origin of the discrepancies between the QCD factorisation predictions for non-leptonic heavy-to-light tree-level $B$-meson decays and the corresponding measurements, further investigations of the QCD factorisation result, and in particular a better understanding 
of the possible size of power corrections, 
is certainly highly desirable.
Moreover, NP effects in non-leptonic tree-level decays would also affect collider observables. A first study \cite{Bordone:2021cca} has found some tension between BSM effects, that would explain the discrepancies in hadronic $B$-meson decays, and dijet resonance searches.
Here, further improvements might have conclusive consequences.
\\
In the case that these tensions persist, it would be instructive to study more 
elaborate scenarios,
in which e.g.\ BSM effects contribute to different $b$-quark transitions. In this regard, 
the channel $b \to c \bar{c} s$ would be particular interesting, since NP in this mode  
could explain the origin of the discrepancies observed in
e.g.\ the $B_s \to \phi \, \ell^+ \ell^-$ decay, 
but it would not affect observables like  $R(K^{(*)})$, see Refs.~\cite{Jager:2019bgk, Jager:2017gal}. 
Furthermore, it would also be worth investigating
explicit 
UV complete BSM models that result in additional contributions
to non-leptonic tree-level decays.
\\
Finally, there is also some potential to improve, from a theoretical point of view, the constraints 
obtained from lifetimes and mixing. Specifically, 
the values of the 
non-perturbative Bag parameters ${\cal B}_{5, \ldots ,8}$ in Eq.~\eqref{eq:Par-ME} could be determined with the HQET sum rules method, following the computations done in Refs.~\cite{Kirk:2017juj, King:2021jsq}, 
while BSM effects from the general effective Hamiltonian in Eq.~(\ref{eq:Heff-NP})
could also be computed
for the dimension-seven four-quark contribution to lifetimes and mixing, as well as for the dimension-six four-quark contribution but at NLO-QCD.
Moreover, as soon as more reliable estimates for the matrix element of the Darwin operator, as well as for the size of the corresponding SU$(3)_F$ breaking effects, 
will be available, it might be interesting to compute BSM contributions also to the coefficients of the dimension-five and dimension-six two-quark operators.

\section*{Acknowledgements}
We thank F.-M. Cai, W.-J. Deng, X.-Q. Li, and Y.-D. Yang for providing us with some of the data from their study 
in Ref.~\cite{Cai:2021mlt}, and 
Matthew Kirk for useful correspondence on the results 
in Refs.~\cite{Kirk:2017juj,King:2021jsq}.
The work of
M.L.P.\ was financed  by the BMBF project {\it Theoretische Methoden für LHCb und Belle II}
(Förderkennzeichen 05H21PSCLA/ErUM-FSP T04).

\appendix

\section{\boldmath Results for the WE diagram with two internal charm quarks}
\label{app:A}
By considering the BSM effective Hamiltonian in Eq.~\eqref{eq:Heff-NP} with the replacement $u \to c$, the contribution from the WE diagram in Figure \ref{fig:PI-WE-NP} to the transition operator in Eq.~\eqref{eq:T-operator} reads
\begin{eqnarray}
{\rm Im}{\cal T}_{\rm NP}^{\rm{WE}, cc} & = & \frac{G_F^2 m_b^2}{6 \pi}|V_{cd}|^2|V_{c b}|^2 \, \sqrt{1 - 4 \rho}
\left[\,\,
\sum_{m, n = 1}^{20} C_m \, C_n^* \, A^{\rm{WE}, cc}_{m n}
- \sum_{m,n = 1}^{2}
C_m^{\rm SM} C_n^{\rm SM} \, A^{{\rm WE},cc}_{mn}\,\,
\right]\,,
\nonumber \\
\label{eq:T-NP-cc}
\end{eqnarray}
with
\begin{eqnarray}
    & \displaystyle
    A^{{\rm WE},cc}_{1,2} = - ((1- \rho) {\cal O}_1 - (1+ 2 \rho) {\cal O}_2)\,, \quad
    A^{{\rm WE},cc}_{1,4} = - 3 \sqrt{\rho} \,  {\cal O}_2\,, \quad 
    A^{{\rm WE},cc}_{1,6} =\frac 3 2 \rho \,  {\cal O}_1\,, 
    &
    \nonumber \\
    & \displaystyle
    A^{{\rm WE},cc}_{1,8} = \frac 3 4 \sqrt{\rho} \, {\cal O}_1\,, \quad
    A^{{\rm WE},cc}_{1,10} = 3 \sqrt{\rho} \,  (4 {\cal O}_2 - {\cal O}_1)\,, \quad 
    A^{{\rm WE},cc}_{1,12} = - 3 \rho \,  {\cal O}_5^\prime \,,
    &
    \nonumber \\
    & \displaystyle
    A^{{\rm WE},cc}_{1,14} = 3 \sqrt{\rho} \,  {\cal O}_6^\prime\,, \quad
    A^{{\rm WE},cc}_{1,16} = \frac12 ((1- \rho) {\cal O}_5^\prime - (1+ 2 \rho) {\cal O}_6^\prime)\,, 
    &
    \nonumber \\
    & \displaystyle
    A^{{\rm WE},cc}_{1,18} =\frac34 \sqrt{\rho}\, ({\cal O}_5^\prime - 2 {\cal O}_6^\prime)\,,
    \quad
    A^{{\rm WE},cc}_{1,20} =-3 \sqrt{\rho} \, ({\cal O}_5^\prime + 2 {\cal O}_6^\prime)\,,
    & 
\end{eqnarray}
\begin{eqnarray}
& \displaystyle
    A^{{\rm WE},cc}_{3,4} = 6 (1- 2 \rho) {\cal O}_2\,, \quad
    A^{{\rm WE},cc}_{3,6} = - \frac 3 2 \sqrt{\rho} \, {\cal O}_2\,, \quad A^{{\rm WE},cc}_{3,8} = - 3 \rho \, {\cal O}_2\,,
    &
    \nonumber \\
    & \displaystyle
    A^{{\rm WE},cc}_{3,10} =- 36 \rho \, {\cal O}_2\,, \quad
    A^{{\rm WE},cc}_{3,14} = - 12 \rho \, {\cal O}_6^\prime  \,, \quad 
    A^{{\rm WE},cc}_{3,16}=\frac 3 2 \sqrt{\rho} \, {\cal O}_6^\prime\,,
    &
    \nonumber \\
    & \displaystyle
    A^{{\rm WE},cc}_{3,18} =\frac32 (1- 2 \rho) {\cal O}_6 \,,\quad 
    A^{{\rm WE},cc}_{3,20} =18 (1- 2 \rho) {\cal O}_6^\prime \,, 
    &
\end{eqnarray}
\begin{eqnarray}
& \displaystyle
    A^{{\rm WE},cc}_{5,6} = - \frac 1 4  ((1 -  \rho) {\cal O}_1 -  (1+ 2 \rho) {\cal O}_2)\,, \quad
    A^{{\rm WE},cc}_{5,8} = \frac 3 8 \sqrt{\rho} \, (2 {\cal O}_2 - {\cal O}_1)\,, \quad 
&   \nonumber \\
&   \displaystyle
    A^{{\rm WE},cc}_{5,10} = 
    \frac32 \sqrt{\rho} \, ({\cal O}_1 + 2 {\cal O}_2)\,, \quad
    A^{{\rm WE},cc}_{5,16} = - \frac 3 4 \rho \, {\cal O}_5^\prime \,, \quad
& \nonumber \\
& \displaystyle
    A^{{\rm WE},cc}_{5,18} = - \frac 3 8 \sqrt{\rho} \, {\cal O}_5^\prime \,, \quad 
    A^{{\rm WE},cc}_{5,20}=\frac32 \sqrt{\rho}\,  ( {\cal O}_5^\prime - 4 {\cal O}_6^\prime ) \,,
    &
\end{eqnarray}
\begin{eqnarray}
& \displaystyle
    A^{{\rm WE},cc}_{7,8} = - \frac 1 8 ((1+ 2 \rho) {\cal O}_1 - 4 (1- \rho) {\cal O}_2) \,, \quad
    A^{{\rm WE},cc}_{7,10} = \frac 1 2 ((1+ 2 \rho) {\cal O}_1 + 4 (2 - 5 \rho){\cal O}_2)\,,
    &
    \nonumber \\
    & \displaystyle
    A^{{\rm WE},cc}_{7,18} = - \frac 3 4  \rho \, {\cal O}_5^\prime\,,
    \quad
    A^{{\rm WE},cc}_{7,20} =3\rho\,({\cal O}_5^\prime - 4 {\cal O}_6^\prime)\,,
    &
\end{eqnarray}
\begin{eqnarray}
& \displaystyle
    A^{{\rm WE},cc}_{9,10} = -2 (1+ 2 \rho){\cal O}_1 + 8 (7 - 13 \rho){\cal O}_2\,, \quad
    A^{{\rm WE},cc}_{9,20} = - 12 \rho \, ({\cal O}_5^\prime + 8 {\cal O}_6^\prime)\,. 
    &
\end{eqnarray}
Again, the remaining functions are obtained from the following replacements:
\begin{equation}
\begin{tabular}{lr}
$\ast \, A^{{\rm WE},cc}_{2m-1, \, 2n-1} = A^{{\rm WE},cc}_{2m - 1, \,2n} 
\Big|_{\Oq^{(\prime)}_i \to \left(\frac{\Oq^{(\prime)}_i}{N_c} + 2 \, \Oq^{(\prime)}_{i+2}\right)}$
&
$m = 1, \ldots, 5\,, \,\,  n = 1, \ldots, 10\,,$
\\[7mm]
$\ast \, A^{{\rm WE},cc}_{2m, \, 2n} = N_c \, A^{{\rm WE},cc}_{2m - 1, \, 2n}\,, \quad A^{{\rm WE},cc}_{2m, \, 2n-1} = A^{{\rm WE},cc}_{2m-1, \, 2n}\,,$ \qquad
& 
$m = 1, \ldots, 5\,, \,\,  n = 1, \ldots, 10\,,$
\\[5mm]
$\ast \, A^{{\rm WE},cc}_{m, \, n} =  A^{{\rm WE},cc}_{n, m}$ \qquad
& 
$
\left\{
\begin{array}{l}
     m = 1, \ldots, 10, \,\, n = \phantom{1}1,  \ldots, 10 \,, \\
     m = 1, \ldots, 10, \,\, n = 11, \ldots, 20 \,, 
\end{array}
\right.
$
\\[6mm]
$\ast \, A^{{\rm WE},cc}_{m, \,n} = A^{{\rm WE},cc}_{m-10, \, n+10} \Big|_{\Oq_i \leftrightarrow \Oqp_i}$
& 
$m = 11, \ldots, 20\,,  n = 1, \, \ldots, 10\,. $
\\[6mm]
$\ast \, A^{{\rm WE},cc}_{m, \,n} = A^{{\rm WE},cc}_{m-10, \, n-10} \Big|_{\Oq_i \leftrightarrow \Oqp_i}$
& 
$m, n = 11, \, \ldots, 20 $\,.
\end{tabular}
\end{equation}
Note that in this case all the entries in the $20 \times 20$ matrix $\hat A^{{\rm WE},cc}$ are non zero.

\section{Supplementary plots}
\label{supplement}
In Figure~\ref{fig:combinations-mW-b-to-c-u-d-only} 
we show the $2\sigma$ contour regions for the NP Wilson coefficients at the scale $\mu = m_W$. Note that, 
since no constraints on paired NP Wilson coefficients for $\mu = m_W$ are given explicitly in Ref.~\cite{Cai:2021mlt}, we do not have bounds
from the non-leptonic $B$-meson decays in this case.
\begin{longfigure}{c}
\includegraphics[width=0.42\textwidth]{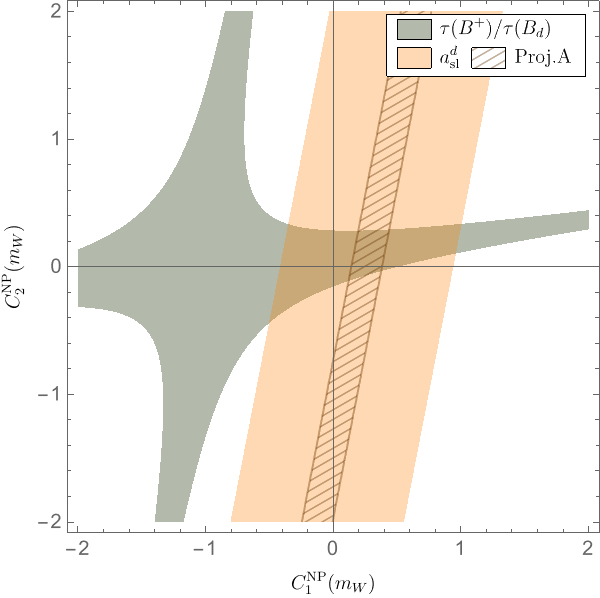}
\qquad 
\includegraphics[width=0.42\textwidth]{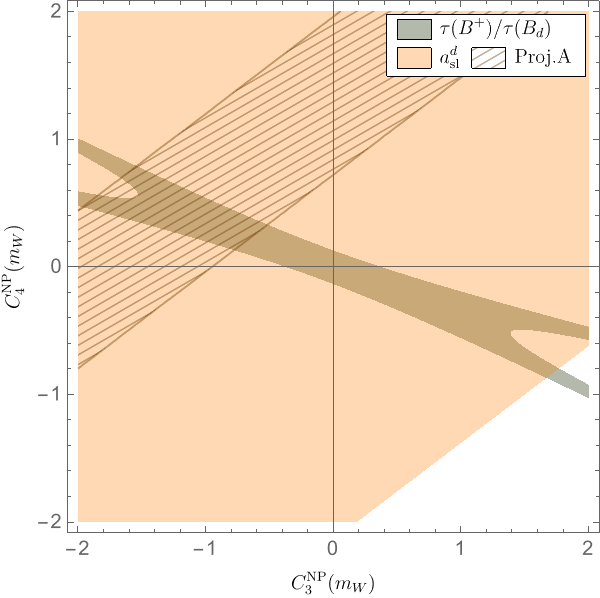} 
\\
\includegraphics[width=0.42\textwidth]{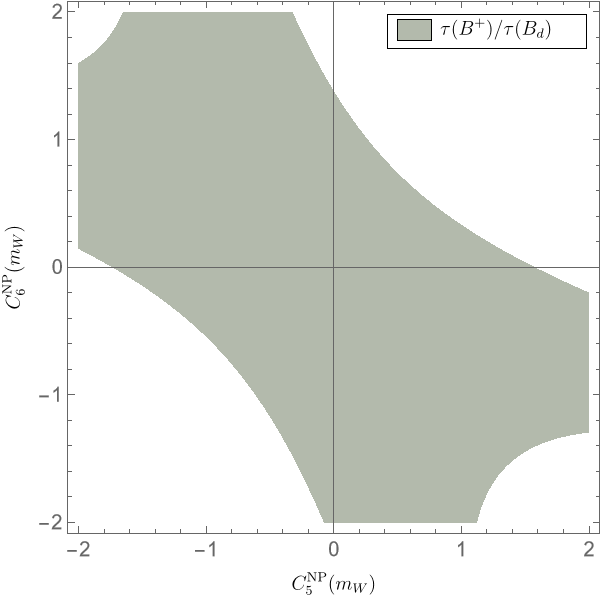}
\qquad 
\includegraphics[width=0.42\textwidth]{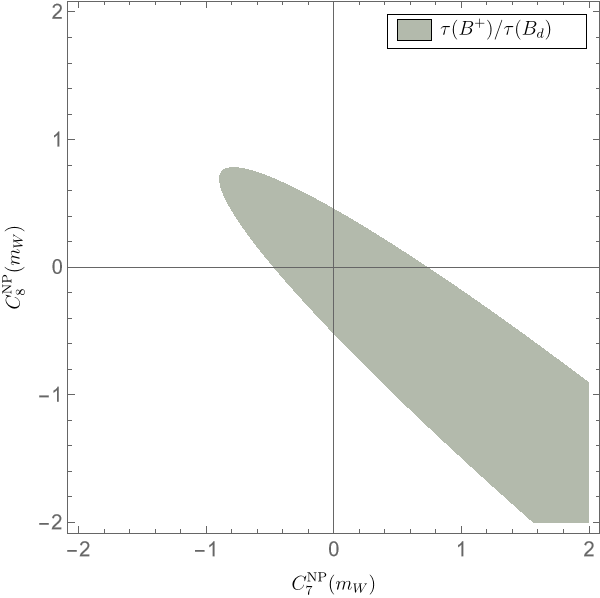} 
\\
\includegraphics[width=0.42\textwidth]{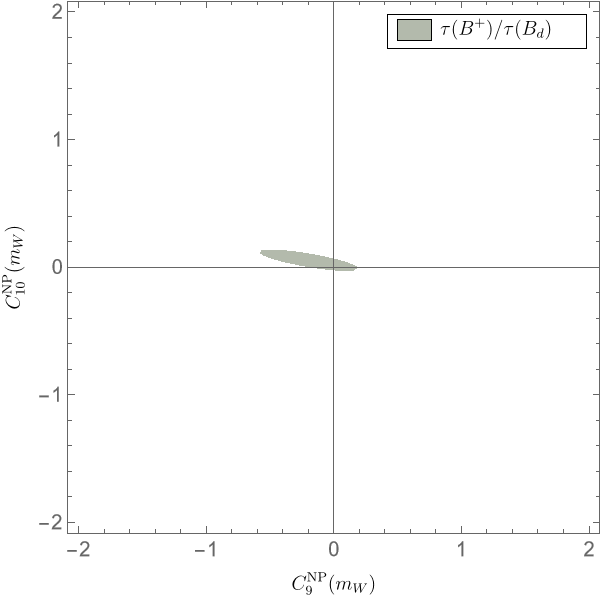}
\qquad 
\includegraphics[width=0.42\textwidth]{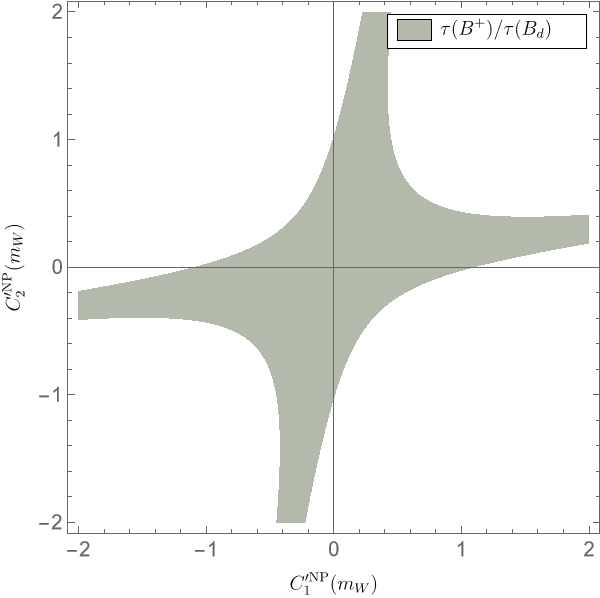} 
\\
\includegraphics[width=0.42\textwidth]{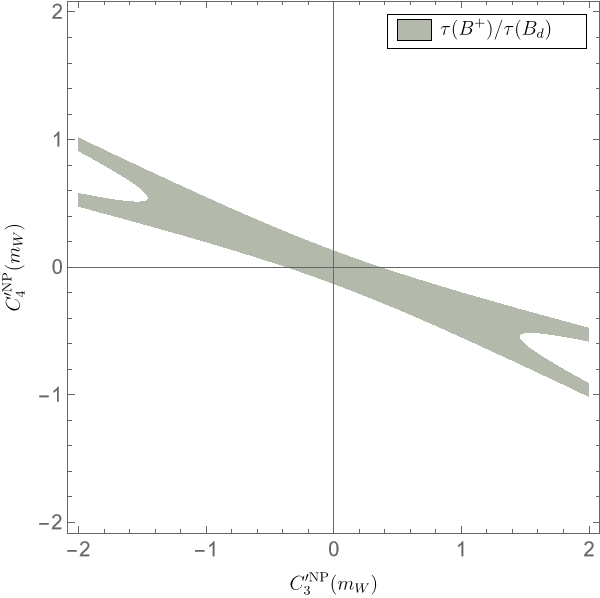}
\qquad 
\includegraphics[width=0.42\textwidth]{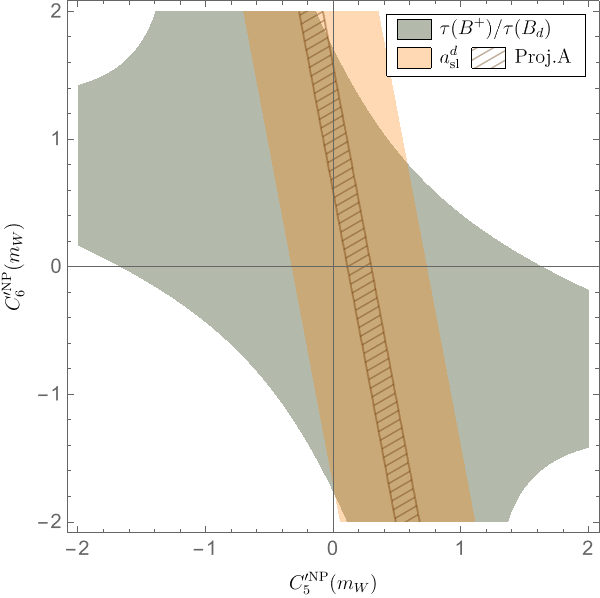} 
\\
\includegraphics[width=0.42\textwidth]{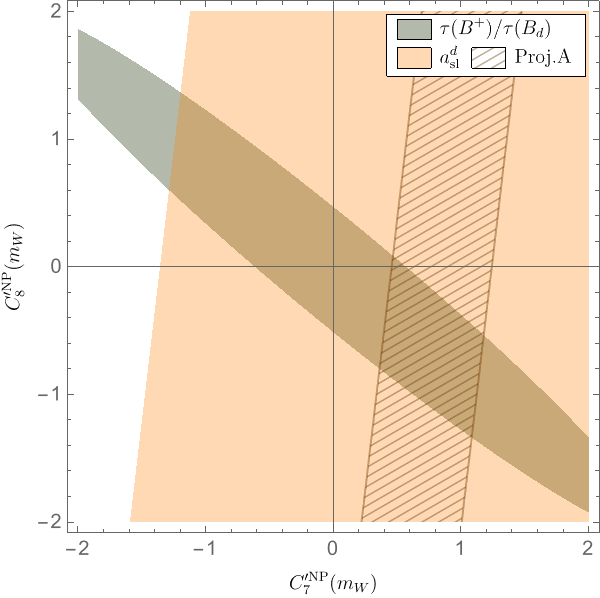}
\qquad 
\includegraphics[width=0.42\textwidth]{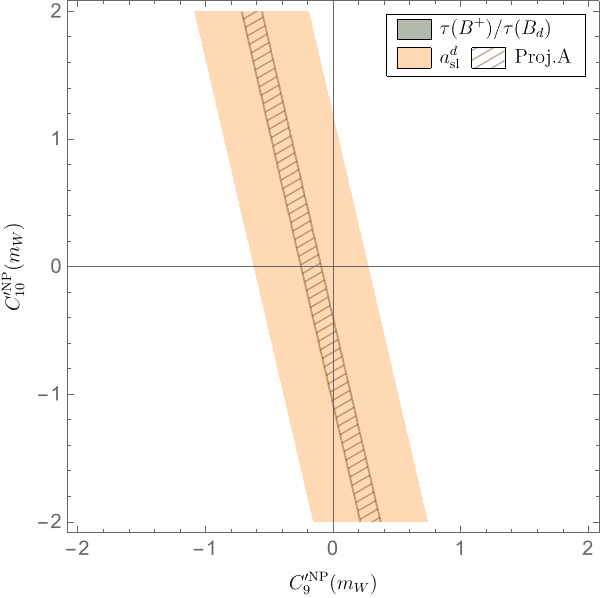} 
\\
\caption{$2\sigma$ contours 
    for the NP Wilson coefficients $C_i^{(\prime) {\rm NP}}(m_W)$, 
    assuming BSM effects only in $b \to c \bar u d (s)$ transitions.}
\label{fig:combinations-mW-b-to-c-u-d-only}
\end{longfigure}

\bibliographystyle{JHEP}
\bibliography{References}

\end{document}